\documentclass[trackchanges,resetfootnote]{aastex701}
\usepackage{amssymb}
\usepackage{placeins}
\usepackage{lipsum}
\usepackage{amsmath, amssymb, amsfonts}
\usepackage{graphicx}         
\usepackage{hyperref}
\usepackage{enumitem}          
\usepackage{bm}               
\usepackage{physics}           
\usepackage{xcolor}            
\usepackage{booktabs}          
\usepackage{tocloft}
\usepackage{subcaption}
\usepackage{comment}
\usepackage{float}
\usepackage{caption}
\usepackage{subcaption}

\begin{document}

\title{Parameterizations of the Hubble Constant from the Binned Type Ia Supernova Master Sample: Logarithmic versus Power-law Forms}

\author[orcid=0000-0003-4442-8546]{Maria Giovanna Dainotti}
\affiliation{Division of Science, National Astronomical Observatory of Japan,
2 Chome-21-1 Osawa, Mitaka, Tokyo 181-8588, Japan}
\affiliation{The Graduate University for Advanced Studies, SOKENDAI, Shonankokusaimura, Hayama, Miura District, Kanagawa, 240-0115, Japan}
\affiliation{Space Science Institute,
4765 Walnut St Ste B, Boulder, CO 80301, USA}
\email[show]{maria.dainotti@nao.ac.jp}

\author[orcid=0009-0009-6648-7249]{Avik Banerjee}
\affiliation{Department of Physics, The University of Burdwan,
Burdwan 713104, West Bengal, India}
\email{avik2020.phys@gmail.com}

\author[orcid=0000-0002-9242-7778]{Andr\'e LeClair}
\affiliation{Physics Department, Cornell University,
Ithaca, NY 14850, USA}
\email{andre.leclair@cornell.edu}

\author[orcid=0000-0002-2550-5553]{Giovanni Montani}
\affiliation{ENEA, Nuclear Department,
C.R. Frascati, Via E. Fermi 45, Frascati 00044, Italy}
\affiliation{Physics Department, Sapienza University of Rome,
P.le A. Moro 5, Rome 00185, Italy}
\email{giovanni.montani@enea.it}

\correspondingauthor{Maria Giovanna Dainotti}

\begin{abstract}

Motivated by the Hubble tension and the increasing debate about the redshift-dependency in the inferred Hubble constant, we investigate its dependence within the flat $\Lambda$CDM framework using a 20-bin analysis of the Master Supernovae type Ia (SNe~Ia) Sample, considering cases with and without very low-redshift data. The main quantity studied is $H_0(z)/H_0$, where $H_0$ is the Hubble constant today, and $H_0(z)$ is the binned data-driven estimate of the value of $H_0$ inferred from the SNe Ia data within redshift intervals at $z > 0$, such that $H_0(0) = H_0$. From the binned analyses, we obtain best-fitting values of $H_0$ and $\Omega_{m0}$, and employ logarithmic and power-law parameterizations, which
are statistically consistent within uncertainties over the redshift range considered.
To assess their behavior at earlier epochs, we extrapolate both forms to the Cosmic Microwave Background radiation (CMB) era ($z\simeq1100$), Big Bang Nucleosynthesis (BBN, $z\sim10^{9}$), and inflationary scales ($z\sim10^{20}$). The reconstructed Hubble constant remains nearly indistinguishable up to the CMB scale, diverges at the few-to-ten percent level around BBN, and differs more substantially when extrapolated to inflationary redshifts, though these two regimes lie beyond the direct observational constraints. A distinct asymptotic behavior emerges at very-high redshift: the logarithmic form exhibits a vanishing behaviour of $H_0$ at finite $z$, while the power-law form approaches zero asymptotically as $z \rightarrow \infty$. In future studies, independent high-redshift observations and extensions beyond $\Lambda$CDM, such as $f(R)$ modified gravity, could allow a comparative study of the two parameterizations beyond the SNe~Ia regime and their high-$z$ physical implications.

\end{abstract}



\section{Introduction}
\begin{figure*}[p]
\centering
\includegraphics[height=0.9\textheight,width=0.85\textwidth]{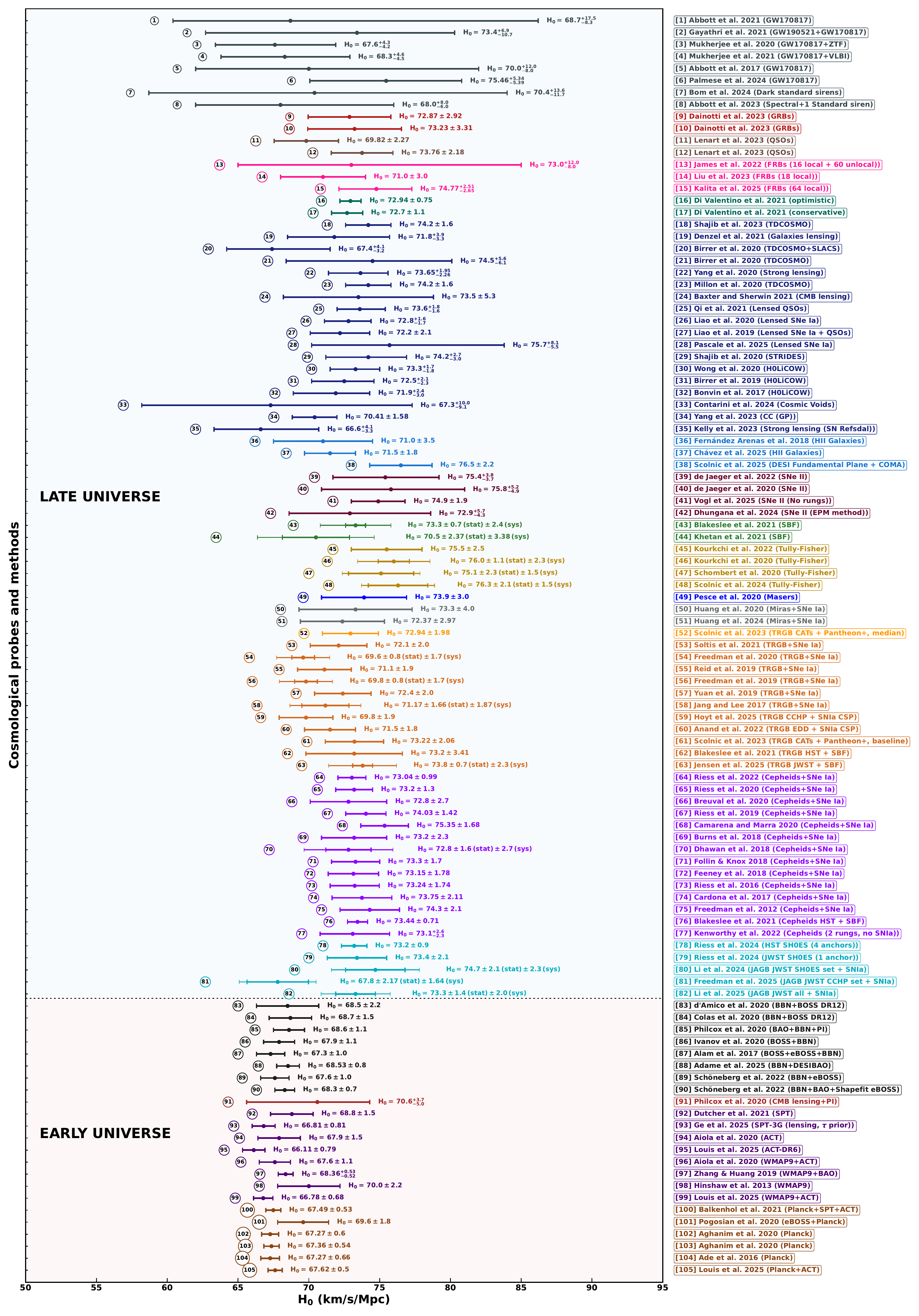}
\caption{
Compilation of $H_0$ measurements from diverse cosmological probes reported in the literature (updated following \citet{2025JHEAp..4800405D} and \citet{2025PDU....4901965D}). Each point represents the central value of a measurement, with the $1\sigma$ uncertainty. For measurements reporting separate statistical (stat) and systematic (sys) uncertainties, the thick inner bars represent the $1\sigma$ statistical errors, while the thin outer bars denote the total $1\sigma$ uncertainties obtained from the quadrature sum of the statistical and systematic components. For clarity, we note that this compilation is intentionally comprehensive and includes a wide range of measurements reported in the literature.
}
\label{fig:H0probes}
\end{figure*}

Over the past few decades, diverse and independent cosmological observations have aligned to support a coherent description of the Universe, establishing the $\Lambda$CDM model \citep{2003RvMP...75..559P} as the standard framework for understanding cosmic expansion and the emergence of large-scale structure \citep{Ostriker:1995su,Bahcall:1999xn,2016A&A...594A..13P,2020A&A...641A...6P,Krauss:1999br}. On large scales, the Universe is well described by the Friedmann–Lemaître–Robertson–Walker (FLRW) metric within general relativity, supplemented by a cosmological constant $\Lambda$ driving late-time acceleration \citep{2001LRR.....4....1C,Peebles:1984ge,2008PASP..120..235R}. This component is commonly associated with dark energy characterized by an equation-of-state parameter $w=-1$, consistent with vacuum energy \citep{2001LRR.....4....1C,1989RvMP...61....1W}, while structure formation arises from the gravitational growth of primordial fluctuations in a cold dark matter (CDM) component \citep{Peebles:1982ff,Blumenthal:1984bp,Davis:1985rj}.

However, the $\Lambda$CDM model faces ongoing challenges on both the theoretical and observational fronts. A key observational issue is the Hubble tension \citep{
2019NatAs...3..891V,
2021CQGra..38o3001D,
2021MNRAS.505.3866E,
2021ApJ...912..150D,
2022PhR...984....1S,
2022JHEAp..34...49A,
2023Univ....9..393V,
2023ARNPS..73..153K,
2023Univ....9...94H,
Dainotti:2023+t,
2025PDU....4901965D,
2025MNRAS.543..456C,
2025PhRvD.112j3515H,
2025PhRvD.112l3509H,
2025EPJC...85.1020B,
2025arXiv250524743M,
2025arXiv250602235Y,
2025arXiv250819770E,
2025arXiv250908840L,
2025arXiv251114332M,
2026A&A...707A.383C,
2026Symm...18..207C,
2026JCAP...02..043B,
2026arXiv260110127L,
2026arXiv260114222S,
2026arXiv260201543H,
2026arXiv260311706K,
Carmesin:2026xua}: a marked discrepancy between local determinations of the current expansion rate and inferences from early-Universe data. The Cepheid-calibrated Type~Ia supernova sample yields a local determination of the Hubble constant of
$H_0 = 73.04 \pm 1.04\ \mathrm{km\ s^{-1}\ Mpc^{-1}}$ \citep{2022ApJ...934L...7R}, while Planck CMB data under $\Lambda$CDM assumptions yields $H_0 = 67.4 \pm 0.5$ km s$^{-1}$ Mpc$^{-1}$ \citep{2020A&A...641A...6P}. The significance of this mismatch varies from $\sim 4\sigma$ to over $6\sigma$, depending on the analysis choices and the sample sizes \citep{2019ApJ...876...85R,2025PDU....4901965D,2020MNRAS.498.1420W,2025arXiv251220193M,2020PhRvR...2a3028C}. 

From a wide range of independent early and late-Universe probes, it is evident that the Hubble tension persists across multiple observational methods. 

In the late Universe, as summarized in \autoref{fig:H0probes}, estimates based on Cepheid-calibrated Type~Ia supernovae \citep[e.g.,][]{2019ApJ...876...85R,2022ApJ...938...36R}, alternative distance-ladder approaches, such as the tip of the red-giant branch \citep[e.g.,][]{2019ApJ...882...34F,2020ApJ...891...57F}, surface-brightness fluctuations \citep{2021ApJ...911...65B}, megamaser galaxies \citep{2019BAAS...51g.176P}, and the Tully--Fisher relation \citep{2020ApJ...896....3K,2022MNRAS.511.6160K} consistently favor higher values of $H_0$. 
\citet{2025ApJ...985..203F} reported updated measurements of the Hubble constant from the Chicago--Carnegie Hubble Program using \textit{JWST} observations, obtaining $H_0$ values based on the Tip of the Red Giant Branch (TRGB) and J-band Asymptotic Giant Branch stars (JAGB) calibrations that are consistent with $\Lambda$CDM expectations and that reduce, though do not eliminate, the discrepancy between early- and late-Universe estimates.
Independent constraints from strong-lensing time-delay measurements \citep[e.g.,][]{2017MNRAS.465.4914B,2020MNRAS.498.1420W,2020A&A...642A.193M}, standard-siren observations from gravitational-wave events \citep{2017ApJ...851L..16A,2021ApJ...909..218A,2025JCAP...07..076C,2026ApJ...997..134Z,2026ApJ..1000..143S,2025arXiv251019931P,2026arXiv260204497G,2025Galax..13..112P} further support this trend. 
Additional probes such as high-redshift probes, including Gamma-Ray Bursts (GRBs) and Quasars \citep{2023MNRAS.518.2201D,2023ApJS..264...46L,2025JHEAp..4700377H,2024JHEAp..44..323F,2026JHEAp..4900439M,2026JHEAp..5100554S} have results tending to higher values of $H_0$. It is of particular interest that an equivalent effective behavior of the redshift density parameter is observed in quasars (see: \citet{2024PDU....4401428D}), and on the Master sample \citep{2025Entrp..27..895M}.

In contrast, early-Universe inferences derived from cosmic microwave background observations by \emph{Planck} \citep{2020A&A...641A...6P}, as well as analyses combining CMB data with baryon acoustic oscillations and Big Bang Nucleosynthesis \citep[e.g.,][]{2017MNRAS.470.2617A,2020JCAP...05..042I}, yield systematically lower values of the Hubble constant, underscoring the persistence of the Hubble tension. \autoref{fig:H0probes} also encompasses a broad compilation of independent measurements of the Hubble constant reported in the literature, spanning a wide range of redshifts and observational data sets. These include results from Cosmic Chronometers (CC), the Atacama Cosmology Telescope (ACT), the South Pole Telescope (SPT), the Carnegie Supernova Project (CSP), surface-brightness fluctuations (SBF), the Extragalactic Distance Database (EDD), the Hubble Space Telescope (HST), the Zwicky Transient Facility (ZTF), H II regions (HII), and fast radio bursts (FRBs), as well as determinations based on the Expanding Photosphere Method (EPM), Gaussian process (GP) reconstructions, and the Comparative Analysis of TRGBs (CATs) approach. This provides a consistent basis for comparing Hubble constant measurements obtained with different observational approaches.

The persistent tension between independent measurements of
$H_0$ has motivated a wide range of alternative theoretical approaches aimed at reconciling these discrepancies. In this context, \citet{2014JCAP...02..002D} show that modifications in the late-time expansion history can lead to a negative contribution from the Integrated Sachs--Wolfe (ISW) effect, suggesting that the observed low-multipole power deficit in the CMB may indicate dark-energy dynamics beyond the standard $\Lambda$CDM model. Among phenomenological extensions of the standard cosmological framework, models allowing for a constant or time-varying dark-energy equation of state, such as the $w$CDM \citep{2003PhRvL..90i1301L} and $w_0w_a$CDM (Chevallier--Polarski--Linder; \citet{2001IJMPD..10..213C,2003PhRvL..90i1301L}) parameterizations, have received considerable attention \citep{2016A&A...594A..13P,2020A&A...641A...6P,2025JCAP...02..021A,2025PhRvD.112h3515A}. In parallel, modified-gravity scenarios \citep{2006CQGra..23.5117S,doi:10.1142/S0219887807001928,2010RvMP...82..451S,2024MNRAS.527L.156M,2025PDU....4801847M,2025PDU....4901969M,2025arXiv251220193M,2025Univ...11..252A,2026arXiv260204559Z,2026arXiv260104048P} have been extensively explored as viable departures from the $\Lambda$CDM paradigm. Interestingly, \citet{2023JCAP...10..047D} discuss the ever present $\Lambda$ cosmological model motivated by causal set theory and unimodular gravity, where the cosmological constant fluctuates in magnitude and sign over cosmic history, offering a potential explanation of the cosmological constant problem without fine tuning.
Among additional proposed interpretations of the Hubble tension, we note that \citet{2025arXiv250202864G} suggested that differences between local and global expansion flows, emerging from structure formation effects within a $\Lambda$-modified weak-field framework, may lead to non-equal effective Hubble parameters.
\citet{2026arXiv260215047C} use the Pantheon+ Type Ia supernova sample and energy-condition constraints to show that $\Lambda$CDM violates the strong energy condition and is disfavored, while the $R_h = ct$ model satisfies all energy conditions and provides a better fit to the data. Furthermore, \citet{2026arXiv260202094S} develop a Lorentz-violating cosmological model based on a spontaneously broken Bumblebee field with Tsallis holographic dark energy that connects early- and late-time expansion and provides an alternative explanation of the Hubble tension.

To quantify potential deviations from $\Lambda$CDM in a model-independent manner, it is useful to introduce an effective Hubble constant as a diagnostic \citep{2021ApJ...912..150D,2022Galax..10...24D, 2021PhRvD.103j3509K,2022arXiv220113384K,2024arXiv240801410S,2025Entrp..27..895M,2026JHEAp..4900459F}:
\begin{equation}
H_0(z) \equiv \frac{H_{\mathrm{model}}(z)}{E_{\Lambda\mathrm{CDM}}(z)} ,
\label{EffectiveHubble}
\end{equation}
where $H_{\mathrm{model}}(z)$ represents the Hubble expansion rate associated with the considered extended cosmological scenario, and $E_{\Lambda\mathrm{CDM}}(z)$ denotes the corresponding dimensionless expansion rate of the $\Lambda$CDM model. 

In recent studies, parametric descriptions \citep{2019MNRAS.483.4803L,2021ApJ...912..150D,2022Galax..10...24D, 2025JHEAp..4800405D,2025arXiv250902636L,2026JHEAp..4900459F} of the Hubble constant, ${H}_0(z)$, or of its effective redshift dependence, have been employed as a phenomenological framework to investigate the physical origin of the Hubble tension. Within this approach, a power-law parametrization of the Hubble constant has been introduced \citep{2021ApJ...912..150D} and shown to provide a viable description in statistical comparisons with a range of theoretical expectations \citep{2025arXiv251116130N,2025arXiv251219568V}. In parallel, scenarios based on the gravitational Casimir effect \citep{2025arXiv250902636L}, which induce an energy scale-dependent effective Newton constant, motivate a logarithmic dependence of the Hubble parameter, as discussed in \citet{2025arXiv250902636L}.

In this work, we perform a 20-bin analysis of the $\Lambda$CDM model to constrain the best-fit cosmological parameters using the Master Sample, a collection of 4 releases (DES5, Pantheon+, Pantheon, and JLA) both including and excluding very low–redshift SNe Ia. This approach enables us to examine the possible redshift dependence of the Hubble constant within the $\Lambda$CDM framework. We revisit the methodologies proposed by \citet{2021ApJ...912..150D,2022Galax..10...24D,2025JHEAp..45..290D} by following the methodology of \cite{2025JHEAp..4800405D}. The statistical assumptions regarding the residuals of SNe Ia distance moduli—normalized by the full covariance matrix—do not satisfy Gaussianity \citep{2024JHEAp..41...30D}. When maximum-likelihood estimators accounting for these non-Gaussian residuals are used (see: \citet{2023MNRAS.521.3909B,2023ApJ...951...63D}), the uncertainties on the inferred cosmological parameters are reduced by up to $\sim 43\%$ \citep{2024JHEAp..41...30D} when also accounting for redshift evolution.
This work stresses the model-independent features of the data and yields a pathway to employ the data analysis of the Master sample to test any model. In this paper, we study two models as a test case and compare them. In the future, any other models can be tested and compared within this data sample. 
Previous studies have investigated the redshift dependence of the cosmic
expansion rate using both parametric and non-parametric approaches, including
Gaussian process reconstructions and alternative binning strategies.
In contrast, the present work focuses on a binned, data-driven estimator of
the Hubble constant constructed from the SNe Ia Master Sample and provides
a direct comparison between logarithmic and power-law parameterizations
within a consistent cosmological framework.
We here fit the Hubble constant by fitting logarithmic \citep{2025arXiv250902636L} and power-law \citep{2021ApJ...912..150D,2022Galax..10...24D} parameterizations of $H_0(z)$ to the binned data. These complementary forms allow us to assess the consistency of the two parameterizations across the redshift range probed and to characterize the nature of any inferred variation. In both cases, we uncover a systematic decrease of the parameterized Hubble constant with increasing redshift \citep{2025JHEAp..45..290D,2025JHEAp..4800405D,2026JHEAp..4900459F}. Finally, motivated by these trends, we extrapolate the best-fit parameterizations to the redshift regimes associated with the CMB, Big Bang Nucleosynthesis, and the inflationary epoch, thereby providing assessments of the early-Universe behavior of the models.

In this work, $H_0$ denotes the present-day Hubble constant, while $H_0(z)$ represents the effective value inferred from binned data at redshift $z$. The normalization obtained from the fit is consistent with $H_0$ within 1 $\sigma$ in the first bin. The paper is organized as follows. \autoref{sec:Theory} outlines the theoretical framework, including the $\Lambda$CDM model and the logarithmic and power-law parameterizations of $H_0$. \autoref{sec:Data} describes the Master Sample. \autoref{model_fitting} presents the methodology, including model fitting and statistical criteria. \autoref{sec:Results} presents the data analysis and results, while \autoref{sec:Discussion} discusses the implications of the findings and their high-redshift extrapolations. \autoref{literature} reviews related work in the context of the current literature. \autoref{sec:Conclusions} summarizes the conclusions. \autoref{sec:AppendixA} derives the statistical $\chi^2$ evaluation and matrix formalism, and \autoref{sec:AppendixB} presents the estimation of $\tilde{H}_0$ from $\chi^2$ minimization. The joint posterior distributions of $H_0$ and $\Omega_{m0}$ for the 20-bin $\Lambda$CDM analysis are presented in \autoref{sec:contours}.
All the Hubble constants and relative quantities are expressed in units of km\,s$^{-1}$\,Mpc$^{-1}$. This unit is used throughout figures and tables unless stated otherwise. 

\section{Framework and parametrization}
\label{sec:Theory}

In this section, we review the standard $\Lambda$CDM model used to compare parameterizations of the Hubble constant using the Master Sample data \citep{2025JHEAp..4800405D}.

In the $\Lambda$CDM framework \citep{2003RvMP...75..559P}, dark energy is modeled as a spatially homogeneous component with constant energy density and negative pressure. It is characterized by a constant equation-of-state parameter $w=-1$.
With this assumption in place, the evolution of the Universe’s expansion is determined by the Friedmann equation, which provides the expression for the Hubble parameter as a function of redshift, $z$ \citep{2003RvMP...75..559P,2025JHEAp..4800405D}:
\begin{equation}
H(z) = H_{0}
\sqrt{
\Omega_{m0}(1+z)^3
+ \Omega_{\Lambda}
+ \Omega_{k0}(1+z)^2
+ \Omega_{r0}(1+z)^4
}.
\end{equation}
Here, $\Omega_{m0}$ and $\Omega_{\Lambda}$ denote the present-day density parameters of non-relativistic matter and the cosmological constant, respectively. The quantity $\Omega_{k0}$ is the density parameter associated with spatial curvature, while $\Omega_{r0}$ represents the radiation density parameter. The radiation component is neglected in the present analysis, as its contribution to the total energy density is negligible over the low-redshift regime considered and does not affect the inferred expansion history or the main objective of this work within the current observational precision.
For a spatially flat Universe, the curvature term vanishes,
$
\Omega_{k0} \simeq 0 ,
$
and all density parameters are evaluated at the present epoch.
Flatness further imposes the condition, $\Omega_{m0} + \Omega_{\Lambda} = 1 $ \citep{peebles1993principles}.

For an expanding Universe described by a homogeneous and isotropic background, the luminosity distance is given by,
\begin{equation}
d_L\!\left(z_{\mathrm{hel}}, z_{\mathrm{HD}}\right)
=
c\,(1+z_{\mathrm{hel}})
\int_{0}^{z_{\mathrm{HD}}}
\frac{dz'}{H(z')} \, .
\end{equation}
Here, $z_{\mathrm{hel}}$ denotes the heliocentric redshift of the supernova, while
$z_{\mathrm{HD}}$ refers to the redshift used in the Hubble diagram after correcting
for the peculiar velocity of the host galaxy. 

For a spatially flat \(\Lambda\)CDM cosmology, the luminosity distance is, 
\begin{equation}
d_L(z_{\rm hel}, z_{\rm HD}) = c \,(1+z_{\rm hel}) 
\int_0^{z_{\rm HD}} \frac{dz'}{H_0 \sqrt{\Omega_{\rm m0} (1+z')^3 + \Omega_\Lambda}} \, .
\end{equation}
This corrected value is computed in
the cosmic microwave background (CMB) rest frame, ensuring that local velocity
perturbations do not bias the inferred cosmological distances (see: \citet{2020ApJ...902...14S,2018ApJ...859..101S,2004PASA...21...97D,2025JHEAp..4800405D}).

Type~Ia supernovae provide precise constraints on the cosmic expansion through measurements of their distance moduli. In cosmological analyses, the observed distance modulus, $\mu_{\mathrm{obs}}$, is compared with its theoretical counterpart, $\mu_{\mathrm{th}}$, which depends on the assumed background cosmology. The theoretical distance modulus (see: \citet{1998AJ....116.1009R,2025JHEAp..4800405D,peebles1993principles}) is defined as:
\begin{equation}
\mu_{\mathrm{th}}
=
5\,\log_{10}\!\left[
\frac{d_L\!\left(z;\Omega_{m0},H_0,\ldots\right)}{\mathrm{Mpc}}
\right]
+ 25 ,
\end{equation}
where $d_L$ represents the luminosity distance measured in megaparsecs. 
\subsection{Parameterization of the Hubble Constant}
\label{Parameterization}
Motivated by the constructions presented in \citet{2021PhRvD.103j3509K,2022arXiv220113384K,2021ApJ...912..150D,2022Galax..10...24D,2025JHEAp..4800405D,2025Entrp..27..895M}, we refer to Eq.~\eqref{EffectiveHubble}
where $H_0(z)$ directly captures departures from the $\Lambda$CDM background evolution. The quantity $H_0(z)$ is a binned, data-driven estimator constructed from redshift intervals of the SNe Ia sample. The functional forms considered below (logarithmic and power-law) provide
data-constrained representations of this estimator and are used to probe
possible redshift-dependent trends in the inferred value of the Hubble constant.

Recent findings \citep{Kazantzidis2020PhRvD.102b3520K,2021ApJ...912..150D,2022Galax..10...24D,2022PhRvD.106d1301O,2024PDU....4401464O,2021PhRvD.103j3509K,2022arXiv220113384K,2023A&A...674A..45J,2024EPJC...84..317M,2025JHEAp..4800405D,2025JHEAp..45..290D,2024arXiv240801410S,2025Entrp..27..895M,2025JCAP...03..026L,2026A&A...707A.189W,2026PhRvD.113f3514D} have unexpectedly shown that a trend of the Hubble constant as a function of redshift is present, independently of the sample considered \citep{2022Galax..10...24D}, the number of bins, and the analysis method \citep{2025JHEAp..4800405D}. This trend should be carefully considered, as ignoring it may lead to a mismatched or incomplete interpretation of the underlying cosmological model. The initial studies, together with subsequent developments, have effectively opened a new research direction that is now being actively investigated by several groups. This behavior has become increasingly intriguing in light of recent observations, including those from DESI.

Motivated by this, we formulate a general parametrized model for the Hubble constant as,
\begin{equation}
\label{H0_model_general}
H_0(z) = \tilde{H}_0 \, g(z; \theta),
\end{equation}
where the parameter $\tilde{H}_0$ is inferred from the fit to the binned data and represents the normalization of $H_0(z)$ at $z=0$. In practice, $\tilde{H}_0$ is consistent with the present-day Hubble constant $H_0$ within uncertainties, although it may depend on the adopted parameterization and dataset.

We stress that $H_0(z)$ should not be confused with the Hubble parameter $H(z) = \dot{a}/a$, which describes the time-dependent expansion rate of the Universe. In contrast, the redshift dependence of $H_0(z)$ arises from the fitting of $H_0$ in each redshift bin.

The function $g(z; \theta)$ encodes the redshift dependence, where $\theta$ denotes the set of model parameters. This formulation provides a unified framework within which different parameterizations can be consistently defined and compared.

In the present work, we examine two representative models, namely the logarithmic \citep{2025arXiv250902636L} and power-law \citep{2021ApJ...912..150D,2022Galax..10...24D,2025JHEAp..4800405D} forms, and assess whether the corresponding values of $\tilde{H}_0$ are consistent with the value of $H_0$ inferred from the total data sample. 
We adopt a common normalization $\tilde{H}_0$ for these parameterizations, corresponding to the present-day value of the Hubble constant, such that 
$H_{0,\log}(0) = H_{0,\mathrm{pl}}(0) = \tilde{H}_0 = H_0$ at $z=0$. In this way, different models describe only the redshift dependence through $g(z;\theta)$, while sharing the same normalization.
            
The theoretical motivations for the logarithmic and power-law parameterizations are discussed in \autoref{2.3} and \autoref{2.4}, respectively. 
\subsection{Quantum vacuum energy density and its renormalization group properties as the origin of the Hubble tension} 
\label{2.3}
\def\bhat{\hat{b}}
\def\rhovac{\rho_{\rm vac}}
\def\mzeron{m_{\rm z}} 
\def\gcal{\mathfrak{g}}
\def\adot{\dot{a}}
\def\addot{\ddot{a}}
\def\gfrak{\gcal}
\def\d{\partial}
\def\inv#1{\dfrac{1}{#1}}
\def\CG{{\cal G}} 
\def\HoCMB{H_{0 ; {\rm CMB}}}
\def\HoSupernova{H_{0 ; {\rm SN}}}
\def\zmax{z_{\rm max}}
\def\amin{a_{\rm min}}
\def\tmin{t_{\rm min}} 

In this subsection, we briefly summarize the theoretical arguments that led to the logarithmic parameterization of the Hubble constant (see: \citet{2025arXiv250902636L}). Within the general framework defined in Eq.~(\ref{H0_model_general}), this corresponds to a specific choice of the function $g(z;\theta)$ given by,

\begin{equation}
\label{H0_log}
H_{0,\log}(z)=\tilde{H}_0
\sqrt{1 - \bhat \ln(1+z)} .
\end{equation}
where $\bhat$ is a free parameter describing the redshift dependence.

The above logarithmic parameterization arose by  exploring the question: 
{\it Can Quantum Vacuum energy be at the origin of Gravity itself?} The  model 
in \citet{2025arXiv250902636L} equates the  cosmologically observed so-called Dark Energy with a cosmological constant 
with equation of state parameter $w=-1$,  where the cosmological constant is attributed  to 
quantum vacuum energy density $\rhovac$.   This aspect of the model is well-founded and prominent in the literature,  however a proper  computation of  $\rhovac$  is very challenging,  since it is expected to be non-perturbative,  and  methods for its computation were not  previously rigorously formulated.           The quantity $\rhovac$  was then studied based on 
the vacuum expectation of the stress-energy tensor:
\begin{equation}
\label{Tvev}
\langle 0 | T_{\mu\nu} |0\rangle = \rhovac \, g_{\mu\nu} , 
\end{equation}
where $T_{\mu\nu}$ is the stress energy tensor for all matter and radiation and $g_{\mu\nu}$ is the flat Minkowski metric.    The above definition of $\rhovac$ guarantees the equation of state parameter 
$w=-1$.    Quantum field theory analysis  led to a specific formula  for $\rhovac$ reviewed below.   Renormalization group  arguments then led to the formula given in Eq.~\eqref{H0_log}.    It should be emphasized that the model considered in \citet{2025arXiv250902636L} should be still viewed as exploratory and is not yet a complete theory.    However, as such it is not a drastic reformulation of General Relativity,  since this just leads to small logarithmic corrections to the Friedmann equations of General Relativity. Since the model is novel,  it is useful to itemize its  primary assumptions.    
 Occam's razor has been honed to an extreme,  and these assumptions are the
following: 

\def\kvec{\bf k} 

\bigskip 
(i)   In the works (\citet{2024JHEP...07..294L,2024JHEP...12..110L}) strong arguments  were  presented  for the following formula,
\begin{equation}
\label{rhovacg}
 \rhovac =\frac{3}{4}  \frac{c^5}{\hbar^3}  \frac{\mzeron^4}{\gfrak}  ,
 \end{equation}
where $\mzeron$ is the physical (renormalized) mass of the {\it lightest}  particle,    and $\gcal$ is a dimensionless coupling constant.   
Two different methods for computing $\rhovac$ were presented which both lead to the 
formula given in Eq.~\eqref{rhovacg}.     The first is based on 
extracting the proper coefficient of the free energy at finite temperature.       The second  was based on 
 the form-factor bootstrap which relates the 2-particle form factor 
$\langle 0 | T_{\mu\nu} |\kvec_1, \kvec_2 \rangle$,  where $|\kvec_1, \kvec_2 \rangle$  is a two particle state with 
wave-vectors $\kvec$, to the 0-particle vacuum expectation value  in  Eq.~\eqref{Tvev}.    This bootstrap approach leads to $\mzeron$ as the physical renormalized mass,   and can explain why $\mzeron$ is the mass of the
lightest particle,   since in principle higher mass particles can be bootstrapped from the lightest.   
For integrable quantum field theories in two spacetime dimensions,   both of these methods lead to the same 
analogous formula for $\rhovac$,   namely $\rhovac \propto \mzeron^2 /\gfrak$, and this formula is known to be {\it exact}.       
  The reason the above formula depends inversely on a coupling $\gfrak$, which is clearly non-perturbative, 
  is easily motivated.  For spacetime dimension 
  $D$ equal to an even integer,  there is an unavoidable logarithmic divergence in $\rhovac$ for {\it free} quantum field theories, namely those with no interaction couplings like $\gfrak$,    that must be regularized such that the $\gfrak \to 0$ limit is  well defined and finite. 
In fact,   this is at the origin of the ``Cosmological Constant Problem",  namely that crude estimates of
$\rhovac$ based on free quantum field theories cannot avoid this divergence such that the ``prediction" is
off  by over a  hundred orders of magnitude,  which was not a solid prediction in the first place. 
The parameter $\mzeron$ sets the scale for $\rhovac$ since $\gfrak$ is dimensionless.  
For the present purposes,   it is not necessary to identify the lightest mass particle for further exploration of the model.    However let us point out that for the current cosmological estimates of $\rhovac$, $\mzeron \approx 0.002 \gfrak^{1/4}$  eV \citep{2025arXiv250902636L}, which is on the order of neutrino masses
\citep{2024PhRvD.110c0001N} 
if   $\gfrak = O(1)$. However, as previously stated the model does not rely on this connection with neutrinos,  although neutrinos are viable candidates for the lightest particle with mass $\mzeron$.

\medskip
(ii)    Arguments were presented in \citet{2025arXiv250902636L}  that the coupling $\gfrak$  should have 
a renormalization group flow in four spacetime dimensions.    The coupling constant  is assumed to be  marginally irrelevant   with the renormalization group beta-function 
 $\mu \d_\mu \gcal =  b \, \gcal^2 /2\pi$ with $b>0$,   where increasing the energy scale $\mu$ corresponds to a flow to higher energies. 
 Integrating this equation, one obtains,
 \begin{equation}
\label{gofmu}
\frac{\gcal (\mu)}{\gcal_0}  = \inv{1 - \bhat    \log (\mu/\mu_0) } ,  ~~~~~ \bhat \equiv  \frac{b \,\gcal_0}{2 \pi}  ,
\end{equation}
where $\gcal_0 \equiv \gcal (\mu_0)$ with $\mu_0$ the energy scale today.  

\medskip
(iii)    There exists an energy scale  $\mu$ of the Universe corresponding to a time dependent  temperature $T(t)$ as for  the $\Lambda$CDM model:
\begin{equation}
\label{Tdef0}
\frac{\mu}{\mu_0}  =  \frac{T(t)}{T_0}  =  \inv{a(t)}  \equiv 1 + z(t) ,
\end{equation}
where $a(t)$ is the scale factor and $T_0$ is the temperature today.   
 
 \bigskip
Based on these assumptions,   consider first a model universe consisting only of vacuum energy density $\rhovac$.   The Friedman equations are the following: 
\begin{equation}
\label{Friedmann}
\frac{\addot}{a} =  \left( \frac{\adot}{a} \right)^2 =   \frac{8 \pi G_N}{3} \rhovac ,
\end{equation}   
where $\rhovac$ should be replaced  by $\rhovac (\mu)$.     
From the expression in Eq.~\eqref{rhovacg} one has,
\begin{equation}
\frac{\rhovac (\mu)}{\rhovac(\mu_0)} = \frac{\gcal_0}{\gcal(\mu)} .
\end{equation}
 In order to express this equation in terms of 
$\rhovac = \rhovac (\mu_0)$ today,   it is   meaningful  to incorporate the renormalization group  flow into an induced flow for Newton's constant $\CG (\mu)$,  
\begin{equation}
\label{FriedmannCG}
  \left( \frac{\adot}{a} \right)^2 =   \frac{8 \pi \CG (\mu) }{3} \rhovac (\mu_0),   
\end{equation}
where,
\begin{equation}
\label{CGdef}
\CG (\mu) =   \CG (\mu_0 )   \frac{\gcal_0}{\gcal (\mu)} =  G_N \cdot   ( 1 - \bhat  \log (\mu/\mu_0) ),  
\end{equation} 
and we have identified $G_N = \CG (\mu_0)$.     For  a marginally irrelevant coupling $\bhat >0$,   note that the effective 
Newton's constant $\CG$ {\it decreases} at higher energy.     
We next add matter and radiation to our model, starting from Eq.~\eqref{FriedmannCG} at a fixed $\mu$.    Consistency with local energy momentum conservation,
namely,   $\nabla^\mu T_{\mu\nu} = 0$,  leads to $\rhovac$ being replaced by the total energy density.
The result can be expressed in the standard form,
\begin{equation}
\label{LCDMHH} 
H_{\log}^2(z) \equiv \left( \frac{\dot{a}}{a} \right)^2  
= \tilde{H}_0^2 \, \bigl(1 + \hat{b} \,\ln a \bigr)
\left(  \frac{\Omega_{\rm rad}}{a^4}  + \frac{\Omega_{\rm m}}{a^3}  + \Omega_\Lambda \right).
\end{equation}
where, by definition, $a(t_0) = 1$ at the present time $t_0$, and $H_{\log}(t_0) = \tilde{H}_0$. Here, $\Omega_\Lambda$ denotes the contribution from the quantum  vacuum energy density $\rhovac$.    
One should emphasize that 
the additional time dependence in the above equation,  due to the pre-factor  $(1 + \hat{b} \ln a(t) )$, does not arise from 
the classical equations of General Relativity,  but rather from the energy scale dependence of the coupling 
$\gfrak$ in 
$\rhovac$ which is tied to the time-dependent temperature $T(t)$.    A useful analogy is that the renormalization group  running of the fine-structure constant in Quantum Electrodynamics does play some role in cosmology so that it increases at higher energies and thus depends on time,   however this additional time dependence is not explicit  in 
Einstein's equations.

The interpretation of the Hubble tension in this model complies with the fitting of the time evolution of $a(t)$ based on 
data which refers to an earlier epoch corresponding to a higher redshift $z$ than today. Then, one should correct for the energy scale-dependent Newton's constant
which led to the factor $(1+ \bhat \log a )$ in Eq.~\eqref{LCDMHH}.      
Without prior knowledge of the trend in $H_0 (z)$ based on Supernovae Ia and proposed  by \citet{2021ApJ...912..150D,2022Galax..10...24D,2025JHEAp..4800405D} and further considered in this article,  in \citet{2025arXiv250902636L}  a rough  estimate of the parameter $\hat{b}$  was given 
 by comparing the value of the Hubble constant inferred from CMB observations, corresponding to $z_{\rm CMB}\simeq1100$, with that obtained from low-redshift supernovae, corresponding to $z_{\rm SN}\simeq0$ (see: \citealt{2020A&A...641A...6P,2022ApJ...934L...7R}),
\begin{equation}
\label{Hos}
\begin{aligned}
 H_{0,\mathrm{CMB}} &= 67.4 \pm 0.5 ~{\rm km\,s^{-1}\,Mpc^{-1}}, \\
 H_{0,\mathrm{SN}} &= 73.0 \pm 1.0 ~{\rm km\,s^{-1}\,Mpc^{-1}}.
\end{aligned}
\end{equation}
Here, $H_{0,\mathrm{CMB}}$ and $H_{0,\mathrm{SN}}$ denote the Hubble constants inferred from the CMB and low-redshift SNe Ia, respectively. Using Eq.~\eqref{H0_log}, we find:
\begin{equation}
\label{HubT2}
\frac{H_{0,\mathrm{CMB}}}{H_{0,\mathrm{SN}}}
=
\frac{\sqrt{1-\hat{b}\ln(1+z_{\rm CMB})}}
{\sqrt{1-\hat{b}\ln(1+z_{\rm SN})}}.
\end{equation}
Substituting the values of $H_{0,\mathrm{CMB}}$ and $H_{0,\mathrm{SN}}$ from Eq.~\eqref{Hos} into Eq.~\eqref{HubT2} for $z_{\rm CMB}=1100$ and $z_{\rm SN}\simeq0$ gives $\hat{b}\approx0.02$. 
Thus, the primary focus of this article is to compare the above rough estimate,  which was originally just based on 2 data points,   to the full trend in $H_0 (z)$ as a function of $z$ for the large data sample of Supernovae used in
\citet{2021ApJ...912..150D,2022Galax..10...24D,2025JHEAp..4800405D}.   This tests the Hubble tension trend and  further constrains the parameter $\hat{b}$ based on a richer cosmological dataset and on an already well-tested binned analysis.

The extra factor $(1+ \hat{b} \log a )$ in Eq.~\eqref{LCDMHH}  significantly modifies the behavior of the early universe, which is already evident in the case of pure vacuum energy.  
To illustrate this,  setting $\Omega_{\rm m} = \Omega_{\rm rad} = 0$, the equation can be explicitly integrated, yielding a modified de Sitter expansion:
\begin{align}
a(t) &= e^{-1/\hat{b}} \,
       \exp\!\left[
       \frac{1}{\hat{b}}
       \left(
       \frac{\hat{b} \tilde{H}_0}{2}(t - t_0) + 1
       \right)^2
       \right] \nonumber \\
     &= \exp\!\left[
       \tilde{H}_0 (t - t_0)
       + \frac{\hat{b} \tilde{H}_0^2}{4}(t - t_0)^2
       \right].
\label{ModdeSit1}
\end{align}
The most interesting feature of the above solution Eq.~\eqref{ModdeSit1} is that there is no curvature singularity $a=0$ for all times.  
 This  minimum value of $a(t)$ is, 
\begin{equation}
\label{amin} 
a(t) > a_{\rm min} \equiv e^{-1/\bhat} ~~\forall t  ~~~~~\Longrightarrow ~~ 
 \zmax = \inv\amin -1 = e^{1/\bhat} -1 .
\end{equation}     
For $\bhat \approx 0.02$,   $\zmax \approx 5 \times 10^{21}$,  which is deep in the radiation dominated era.   
Let us emphasize that the model does not break down at $\zmax$,   which can be attributed to the fact that the effective Newton's constant 
vanishes at this $z$ and the above solution is valid at times before when $a(t) = \amin$.        This can be seen from the explicit solution 
Eq.~\eqref{ModdeSit1}  which  has the remarkable symmetry,
\begin{equation}
\label{minustsymmetry}
a(t) = a(-t + 2 \tmin) ,
\end{equation}
where $a(\tmin) = \amin$,  which implies the solution extends to before the time $\tmin$.    
The above symmetry is independent of the present time $t_0$.
It was argued in \citet{2025arXiv250902636L} that these features of a minimal scale factor $\amin$ and maximum redshift $\zmax$ persist upon the addition of matter and 
radiation, in particular for the $\Lambda$CDM model.      A hot Big Bang can be associated with the time $\tmin$ where the universe is at its hottest,   and the symmetry  Eq.~\eqref{minustsymmetry} opens  
the opportunity  to address what occurred before this hot Big Bang.  
\subsection{Theoretical Realizations of the Power-Law behavior of the effective Hubble Constant}
\label{2.4}
The power-law behavior of the effective Hubble constant, introduced in \citet{2021ApJ...912..150D}, resulted in being the most favoured profile by statistical comparisons with predictions from various theoretical formulations; see, for instance, \citet{2026JHEAp..4900459F,2025arXiv251116130N,2025arXiv251219568V}. 
The power-law parameterization of the effective Hubble constant, $H_{0,\mathrm{pl}}(z)$, discussed in \citet{2021ApJ...912..150D,2022Galax..10...24D,2025JHEAp..4800405D}, is expressed as:
\begin{equation}
\label{2.4Eq}
H_{0,\mathrm{pl}}(z)=\frac{\tilde{H}_0}{
(1+z)^{\alpha}},
\end{equation}
where $\alpha$ is the evolutionary coefficient governing the redshift dependence. 
Following the same approach as in \autoref{2.3} and \citet{2025arXiv250902636L}, we estimate the parameter $\alpha$ by relating the CMB scale ($z_{\rm CMB}\simeq1100$) to the low-redshift supernova scale ($z_{\rm SN}\simeq0$). Eq.~\eqref{2.4Eq} then gives,
\begin{equation}
\label{PLValue}
\frac{H_{0,\mathrm{CMB}}}
     {H_{0,\mathrm{SN}}}
=
\frac{(1+z_{\rm CMB})^{-\alpha}}
     {(1+z_{\rm SN})^{-\alpha}}.
\end{equation}
Substituting the values of $H_{0,\mathrm{CMB}}$ and $H_{0,\mathrm{SN}}$ from Eq.~\eqref{Hos} into Eq.~\eqref{PLValue} for $z_{\rm CMB}=1100$ and $z_{\rm SN}\simeq0$ gives:
$\alpha \approx 0.01.$

Due to its phenomenological formulation, related to typical scaling laws in the redshift evolution of astrophysical sources \citep{2024ApJ...963L..12P}, we could be led to infer that this preference in the data analysis is a possible indication of a hidden redshift evolution of the parameters involved in SNe Ia. This possibility is open to scientific debate, but, as far as a direct indication emerges from the data, we have to consider different points of view, mainly related to modifications of the late Universe dynamics with respect to a $\Lambda$CDM-model \citep{2024PDU....4401486M,2025PDU....4801847M,2025PDU....4901969M}. In this respect, models able to reproduce the power-law behavior of the effective Hubble constant stand for their relevance and deserve detailed investigation, as done in the study of \citet{2025arXiv250902636L}.
In fact, we have shown that, for sufficiently small values of the redshift, the discussed gravitational Casimir effect \citep{2025arXiv250902636L}, responsible for an effective changing Newton constant, provides exactly the power-law scaling as soon as the identification $\hat{b}=2\alpha$ between the two parameters is implemented (see: \autoref{sec:2.5}).

Another important representation of the power-law decaying behavior of the effective Hubble constant has been achieved in \citet{2023MNRAS.522L..72S}, where a metric $f(R)$-gravity has been implemented, in the so-called Jordan frame \citep{2010RvMP...82..451S}, to the cosmological problem. \citet{2023MNRAS.522L..72S} demonstrated that the redshift dependence of the Hubble constant, inferred from the Pantheon SN~Ia sample may be interpreted as an effective rescaling of the Einstein constant induced by a nearly constant scalar-field potential for $z \lesssim 0.3$, yielding an $f(R)$ model consistent with the observed trend. \footnote{A different formulation of the same physics was proposed in \citet{2024MNRAS.527L.156M} where the power-law scaling has been exactly reproduced in a viable modified gravity scenario, and the obtained picture is \emph{de facto} re-interpreted as a rescaling of the Einstein constant, according to the original ideas stated in \citet{2021ApJ...912..150D,2022Galax..10...24D}.} Thus, the analysis in \citet{2023MNRAS.522L..72S}, and \citet{2024arXiv240801410S}, has qualitatively the same final prediction as the study in \citet{2025arXiv250902636L}, which we deepened above. Within this framework, \citet{2024PDU....4401486M} constructed a fully redshift-dependent scalar-field dynamics capable of varying Hubble constant while preserving consistency with both local and high-redshift measurements; the resulting profile matched the SN~Ia value at $z=0$ and converged to $\Lambda$CDM at higher redshifts without compressing the supernova information.

A different physical situation, able to reproduce very well the power-law profile, has been achieved in \citet{2025Entrp..27..895M}, where an interaction model between dark energy and dark matter is formulated on a phenomenological level, assuming that the former is subject to a process of constituent creation by the gravitational field of the expanding Universe. This study is of relevance because, although the comparison with the power-law behavior is performed numerically only, it clarifies that such a rescaling of the observed Hubble constant can also be described by a deeper physical picture from a dynamical point of view with respect to the simple scaling of Newton’s or Einstein’s constant, which, however, still remains the most natural solution to explain the observed profile.

In \citet{2025arXiv250713890F}, inflation is investigated within the $f(R)$ gravity metric in the Jordan frame, where the effective scalar degree of freedom drives a slow--roll, quasi--de Sitter phase, smoothly matches $\Lambda$CDM after inflation, incorporates radiation--like particle production, and is tested against Pantheon+ and DESI data with implications for the Hubble constant tension. Building on these results, \citet{2025PDU....4901969M} investigated a late--time cosmological scenario, where dark energy decayed into dark matter and gave rise to an effective redshift-dependent Hubble parameter; comparison with binned Pantheon Supernova Ia data constrained the additional model parameter, yielded an improved low-redshift fit relative to the power-law phenomenology, and only weakly impacted the Hubble constant tension without extension to recombination. Similarly, \cite{2025arXiv251116130N} showed that a viscous dark energy component generated by the Hubble flow can reproduce the effective redshift dependence of the Hubble constant inferred from the binned Master Sample, while \cite{2025arXiv251219568V} investigated another dynamical cosmological framework capable of reproducing the observed evolution of $H_0(z)$ using the same redshift-binned Master Sample analysis.

Finally, we would like to stress that the possibility of a changing Einstein constant with the Universe’s energy density has been investigated in \citet{2025EPJC...85..881M}, where the validity of the Bianchi identities has been preserved. This scenario provided a non-zero vacuum energy density for the Universe’s asymptotic evolution, associated with an anomalous pressure contribution for the dark energy density term. In the end, there emerges a (non-conformal but additive) deviation from the standard $\Lambda$CDM model via a logarithmic term, as the one obtained in \citet{2025arXiv250902636L}. The comparison of the emerging modified model with all the low-redshift sources provided an indication that the parameter multiplying the logarithmic contribution is compatible with zero. Regardless, it remains an interesting perspective, on the one hand there is the theoretical comparison of these two models in \citet{2025arXiv250902636L} and \citet{2025EPJC...85..881M}, and, on the other hand, the implementation of the latter analysis in the prediction of a varying Hubble constant and the evaluation of the parameter controlling the logarithmic term as determined via the binned data of the SN Ia Master sample \citep{2025JHEAp..4800405D}.

\subsection{Comparing the Logarithmic and Power-Law Parameterizations of the Hubble Constant}
\label{comparison}
We adopt the parametrization introduced in \autoref{2.3} to describe the redshift dependence of the Hubble constant. We consider a logarithmic parameterization \citep{2025arXiv250902636L} where $\tilde{H}_0$ and $\hat{b}$ in Eq.~\eqref{H0_log} are determined using a weighted least-squares fitting procedure \citep{bevington2003data,Press:2007ipz}.

In the local cosmological regime, where $z$ is sufficiently small, the quantity $\lvert \hat{b}\,\ln(1+z) \rvert$ becomes small. Expanding Eq.~\eqref{H0_log} in a Taylor series, we obtain,
\begin{equation}
H_{0,\log}(z)
\simeq
\tilde{H}_0
\Big[
1 - \frac{\hat{b}}{2}\ln(1+z)
- \frac{\hat{b}^{2}}{8}\big(\ln(1+z)\big)^{2}
- \frac{\hat{b}^{3}}{16}\big(\ln(1+z)\big)^{3}
+ \cdots
\Big] .
\label{Log_expansion}
\end{equation}

We next consider a power-law parameterization \citep{2021ApJ...912..150D,2022Galax..10...24D,2025JHEAp..4800405D}, given by Eq.~\eqref{2.4Eq}, that can equivalently be written in exponential form as,
\begin{equation}
H_{0,\mathrm{pl}}(z)
=
\tilde{H}_0\,
\exp\!\big[-\alpha\,\ln(1+z)\big].
\label{PL_exp}
\end{equation}

In the small-$z$ regime, Eq.~\eqref{PL_exp} can be expanded in a Taylor series, yielding,
\begin{equation}
H_{0,\mathrm{pl}}(z)
\simeq
\tilde{H}_0
\Big[
1
- \alpha\ln(1+z)
+ \frac{\alpha^{2}}{2}\big(\ln(1+z)\big)^{2}
- \frac{\alpha^{3}}{6}\big(\ln(1+z)\big)^{3}
+ \cdots
\Big] .
\label{PL_expansion}
\end{equation}
A comparative analysis between the two parameterizations enables us to assess the redshift interval over which they yield consistent predictions and where they begin to deviate.
\section{The Data Sample}
\label{sec:Data}

In this work, we adopt the Master Sample of Type Ia supernovae presented in \citet{2025JHEAp..4800405D}. This dataset was constructed through the combination of four major SNe Ia catalogs: the Dark Energy Survey Supernova Program (DES; \citet{2024ApJ...973L..14D}), the Pantheon+ compilation (P+; \citet{2022ApJ...938..113S,Brout_2022}), the Pantheon compilation (\citet{2018ApJ...859..101S}), and the Joint Light-curve Analysis (JLA; \citet{2014A&A...568A..22B}), with a careful identification and removal of duplicate events. The resulting compilation contains 3714 unique SNe Ia and is hereafter referred to as the \emph{Master Sample}. The merging procedure implemented by \citet{2025JHEAp..4800405D} assigns priority to surveys in reversed chronological order (DES, P+, Pantheon, and JLA), ensuring that each supernova appears only once in the final dataset.
The composition of the Master Sample, as reported in \citet{2025JHEAp..4800405D}, is summarized in \autoref{tab:master_sample}. 

The Master Sample is constructed within a statistical framework that relaxes the assumption of Gaussian likelihoods, motivated by evidence that the covariance-normalized residuals of SNe~Ia distance moduli exhibit non-Gaussian features (\citet{2024JHEAp..41...30D,2025MNRAS.536..234L}). The use of optimized likelihood functions (\citet{2023MNRAS.521.3909B,2023ApJ...951...63D}) has been shown to influence the inferred uncertainties on cosmological parameters, while the associated redshift-binning procedure enables the examination of possible redshift-dependent trends in the data, which may arise from astrophysical systematics or departures from standard cosmological assumptions. 
We have already verified that the main results are robust against reasonable variations
in the binning scheme, including changes in the number of bins and binning
strategy, with no significant impact on the inferred trends, thus indicating that the inferred trends are not driven
by binning choices, see \citep{2025JHEAp..4800405D}.

For the full Master Sample, the Hubble-diagram redshift,
denoted by $z_{\mathrm{HD}}$, spans the range
$0.00122 \leq z_{\mathrm{HD}} \leq 2.26137$. In the present analysis, we consider two variants of this dataset.
The first consists of the full Master Sample, covering the entire redshift interval above,
and is hereafter referred to as the \emph{Master Sample with low-$z$}. The second variant excludes very low-redshift supernovae by imposing a lower cut at
$z_{\mathrm{HD}} = 0.01006$, such that
$0.01006 \leq z_{\mathrm{HD}} \leq 2.26137$.
This reduced dataset is hereafter referred to as the \emph{Master Sample without low-$z$}.
The exclusion of very low-redshift SNe~Ia allows us to assess the impact of peculiar velocities \citep{2006PhRvD..73l3526H,2011ApJ...741...67D,2022ApJ...938..112P}, which can significantly affect distance measurements in the nearby Universe. Each of the two datasets is independently divided into 20 redshift bins. The binning scheme used here is the equi-populated approach, in which each bin contains approximately the same number of SNe~Ia, ensuring a sufficient number of data points in every redshift bin to obtain statistically reliable constraints, following the approach of \citet{2021ApJ...912..150D}. In \citet{2025JHEAp..4800405D}, additional binning schemes, namely the moving-window (MW) and logarithmic-$z$ equi-spacing approaches, were also considered, and the decreasing trend in $H_0(z)$ was found to be quantitatively consistent (within $1\sigma$) across all three methods. Therefore, any of these three methods are equivalent and here we adopts the equi-populated binning.
A uniform prior $\mathcal{U}(60,80)$ was imposed on the Hubble constant $H_{0}$, while the matter density parameter $\Omega_{\mathrm{m}0}$ was constrained by a Gaussian prior $\mathcal{N}(0.322,\,0.025)$, corresponding to the mean and the standard deviation obtained from a
$\Lambda$CDM fit to the full dataset.

\begin{table*}[t]
\centering
\caption{Composition of the Master Sample \citep{2025JHEAp..4800405D}}
\label{tab:master_sample}
\resizebox{0.6\textwidth}{!}{%
\begin{tabular}{lccc}
\hline
\textbf{Dataset} & \textbf{Original SNe Ia} & \textbf{Duplicates Removed} & \textbf{Final Contribution} \\
\hline
DES      & 1829 & 0   & 1829 \\
Pantheon+       & 1701 & 493 & 1208 \\
Pantheon & 1048 & 867 & 181  \\
JLA      & 740  & 244 & 496  \\
\hline
\textbf{Total} &  &  & \textbf{3714} \\
\hline
\end{tabular}
}
\end{table*}

Following the formulation adopted in the Master Sample, the observed distance modulus,
$\mu_{\mathrm{obs}}$, is computed using a modified Tripp relation \citep{1998A&A...331..815T}, which links
Type~Ia supernova photometric observables to standardized luminosities. The expression can
be written as: 
\begin{equation}
\mu_{\mathrm{obs}}
=
m_{B}
-
M_{B}^{0}
+
\eta_{S}\, S^{\mathrm{LC}}
-
\eta_{C}\, C^{\mathrm{SN}}
+
\Delta_{\mathrm{host}}
+
\Delta_{\mathrm{bias}} ,
\end{equation}
where $m_{B}$ is the observed rest--frame peak magnitude in the $B$ band, and $M_{B}^{0}$
denotes the absolute magnitude of a reference supernova corresponding to
$S^{\mathrm{LC}} = 0$ and $C^{\mathrm{SN}} = 0$. The quantities $S^{\mathrm{LC}}$ and $C^{\mathrm{SN}}$
represent the light--curve stretch and color parameters, respectively, while
$\eta_{S}$ and $\eta_{C}$ are the free parameters. The term
$\Delta_{\mathrm{host}}$ accounts for empirical corrections related to the host--galaxy
stellar mass, whereas $\Delta_{\mathrm{bias}}$ incorporates bias corrections derived from
survey simulations \citep{2018ApJ...859..101S}. As discussed in \citet{1998A&A...331..815T} and \citet{2018ApJ...859..101S}, Type~Ia supernova observations exhibit a degeneracy between the absolute magnitude and the Hubble constant (see: \citet{2025JHEAp..4800405D}). 

This degeneracy is removed by calibrating the absolute magnitude $M_{B}^{0}$ using a reference value $H_{0}=70\,\mathrm{km\,s^{-1}\,Mpc^{-1}}$, which fixes the distance normalization without affecting the behavior of the effective Hubble constant.
\section{Methodology: Model Fitting and Statistical Criteria} 
\label{model_fitting}
To determine the best-fit values of the model parameters, we employ the method of least-squares fitting. In this approach, the parameter estimation is performed by minimizing the chi-square ($\chi^{2}$) statistic \citep{bevington2003data}. In addition to the least-squares analysis, we perform model comparison using the Bayesian Information Criterion (BIC; \citet{2007MNRAS.377L..74L}). By analyzing the results obtained from each binned dataset, we assess the relative performance of the two models and identify the parameterization that provides a statistically preferred description of the data.

For both the logarithmic parametrization given in Eq.~\eqref{H0_log} and the power law in Eq.~\eqref{2.4Eq}, the associated $\chi^{2}$ statistic is expressed generically as:
\begin{align}
\label{chi2_Log}
\chi^2_{model}
&=
\sum_{i=1}^{20}
\left[
\frac{
H_0(z_i)
-
\tilde{H}_0\, g(z_i; \theta)
}{
\sigma_{H_0(z_i)}
}
\right]^2 .
\end{align}
where $H_0(z_i)$ is the best-fit value of $H_0$ in the $i$-th redshift bin of the Master Sample, obtained by independently fitting a flat $\Lambda$CDM model using a Markov Chain Monte Carlo (MCMC) analysis, see \autoref{sec:Results}, $\sigma_{H_0(z_i)}$ represents the uncertainty associated with the obtained Hubble constant, $H_{0}(z_i)$ in the $i$-th redshift bin centered at $z_i$. The summation is performed over the full set of $20$ bins, with $i = 1, 2, \ldots, 20$.\\
We write the $\chi^{2}$ statistic for both models in matrix form as (see: \autoref{sec:AppendixA}; \citet{bevington2003data,Press:2007ipz}), 
\begin{align}
\label{chi2_mat}
\chi^2_{model}
&=
\left[
\Delta H_0^{model}\!\left(\tilde{H}_0,\theta\right)
\right]^{T}
C_{H_0}^{-1}
\left[
\Delta H_0^{model}\!\left(\tilde{H}_0,\theta\right)
\right] .
\end{align}
The covariance matrix ${C_{H_0}}$ corresponding to the Hubble constant measurements is given by:
\begin{equation}
{C_{H_0}} = \mathrm{diag}\,\Big( \sigma_{H_0(z_1)}^2,\, \sigma_{H_0(z_2)}^2,\, \ldots,\, \sigma_{H_0(z_n)}^2 \Big)\,.
\label{Cmatrix}
\end{equation}
We use this assumption for the Covariance matrix, since inter-bin correlations are negligible because $H_0$ values are independently inferred within each bin, considering each bin as a distinct SN subset.

\subsection{Equivalence condition for logarithmic and power-law parametrizations of the Hubble constant}
\label{sec:2.5}
For convenience, we define the following functions,
\begin{equation}
\label{Eq.34}
g_{\mathrm{\log}}(z_i;\hat{b})
\equiv
\sqrt{1-\hat{b}\ln(1+z_i)} .
\end{equation}
\begin{equation}
\label{Eq.39}
g_{\mathrm{pl}}(z_i;\alpha)
\equiv
(1+z_i)^{-\alpha} 
\\
=
\exp\!\big[-\alpha \ln(1+z_i)\big].
\end{equation}

We now perform a Taylor expansion of both functions, 
$g_{\text{log}}(z_i;\hat{b})$ and $g_{\text{pl}}(z_i;\alpha)$, given in Eqs.~\eqref{Eq.34} and \eqref{Eq.39}, respectively, and we obtain:
\begin{equation}
g_{\mathrm{\log}}(z_i;\hat{b})
\simeq
1
- \frac{\hat{b}}{2}\ln(1+z_i)
- \frac{\hat{b}^{2}}{8}\big(\ln(1+z_i)\big)^{2}
- \frac{\hat{b}^{3}}{16}\big(\ln(1+z_i)\big)^{3}
+ \cdots
\label{gLog_expansion}
\end{equation}
Here, the quantity $|\hat{b}\,\ln(1+z_i)|$ is assumed to be sufficiently small.

\begin{equation}
g_{\mathrm{pl}}(z_i;\alpha)
\simeq
1
- \alpha \ln(1+z_i)
+ \frac{\alpha^{2}}{2}\big(\ln(1+z_i)\big)^{2}
- \frac{\alpha^{3}}{6}\big(\ln(1+z_i)\big)^{3}
+ \cdots
\label{gPL_expansion}
\end{equation}

In the local cosmological regime, where the redshift $z_i$ is sufficiently small,
we keep only the leading linear contribution in $\ln(1+z_i)$ and neglect all
higher-order terms.\\
Therefore, using Eqs.~\eqref{gLog_expansion}, and \eqref{gPL_expansion}, in the limit of small redshift,
\begin{equation}
\begin{aligned}
    g_{\text{log}}(z_i; \hat{b}) &\simeq 1 - \frac{\hat{b}}{2}\,\ln(1+z_i) \\
    g_{\text{pl}}(z_i; \alpha) &\simeq 1 - \alpha\,\ln(1+z_i)
\end{aligned}
\label{eq:smallz_g_equivalence}
\end{equation}
When $\hat{b} = 2\alpha$, the two functions satisfy,
\begin{align}
g_{\mathrm{\log}}(z_i;\hat{b})
&\approx
g_{\mathrm{pl}}(z_i;\alpha).
\end{align}
From the two parameterizations, given in Eqs.~\eqref{H0_log}, and \eqref{2.4Eq}, we may write,
\begin{equation}
\label{eq.32}
\begin{aligned}
H_{0,\log}(z_i) &= \tilde{H}_0\, g_{\log}(z_i; \hat{b}) \\
H_{0,\mathrm{pl}}(z_i)  &= \tilde{H}_0\, g_{\mathrm{pl}}(z_i; \alpha)
\end{aligned}
\end{equation}
Thus, in the small-$z$ regime, when $\hat{b} = 2\alpha$, we have,
\begin{equation}
H_{0,\log}(z_i) \simeq H_{0,\mathrm{pl}}(z_i).
\end{equation}

Therefore, in the small-redshift regime where $\hat{b} = 2\alpha$, the logarithmic and power-law parameterizations of the Hubble constant become equivalent at leading order.

\section{Data Analysis and Results}
\label{sec:Results}
As discussed in \autoref{sec:Data}, the Master SNe Ia Sample, both including and excluding low-redshift SNe Ia, is divided into 20 redshift bins equally populated. For the $\Lambda$CDM cosmological model, parameter inference is performed independently for each redshift bin in both datasets using MCMC sampling implemented through the \texttt{Cobaya} framework
\citep{2021JCAP...05..057T}, to compare theoretical predictions with observational measurements. 
The convergence of the chains is assessed using the Gelman--Rubin diagnostic
\citep{1992StaSc...7..457G}, requiring $R-1 < 0.01$. Posterior distributions and derived
constraints are analyzed and visualized with the \texttt{getdist} package \citep{2025JCAP...08..025L}. Flat priors are assumed for the cosmological parameters, with
$\Omega_{\mathrm{m}0} \sim \mathcal{U}(0.01,\,0.99)$ and
$H_{0} \sim \mathcal{U}(60,\,80)\,\mathrm{km\,s^{-1}\,Mpc^{-1}}$.
For each bin, the joint posterior distributions of $H_{0}$ and $\Omega_{\mathrm{m}0}$ for both data
selections are derived, for details see \autoref{sec:contours}. The inner and outer contours represent
the $1\sigma$ (68\%) and $2\sigma$ (95\%) confidence regions, respectively. The resulting set of 20 best-fit $H_0$ values constitutes the observational $H_{0}(z_i)$ dataset used in the subsequent analysis.

\begin{figure}[p]
    \centering

    \begin{minipage}{0.46\textwidth}
        \centering
        \includegraphics[width=\linewidth]{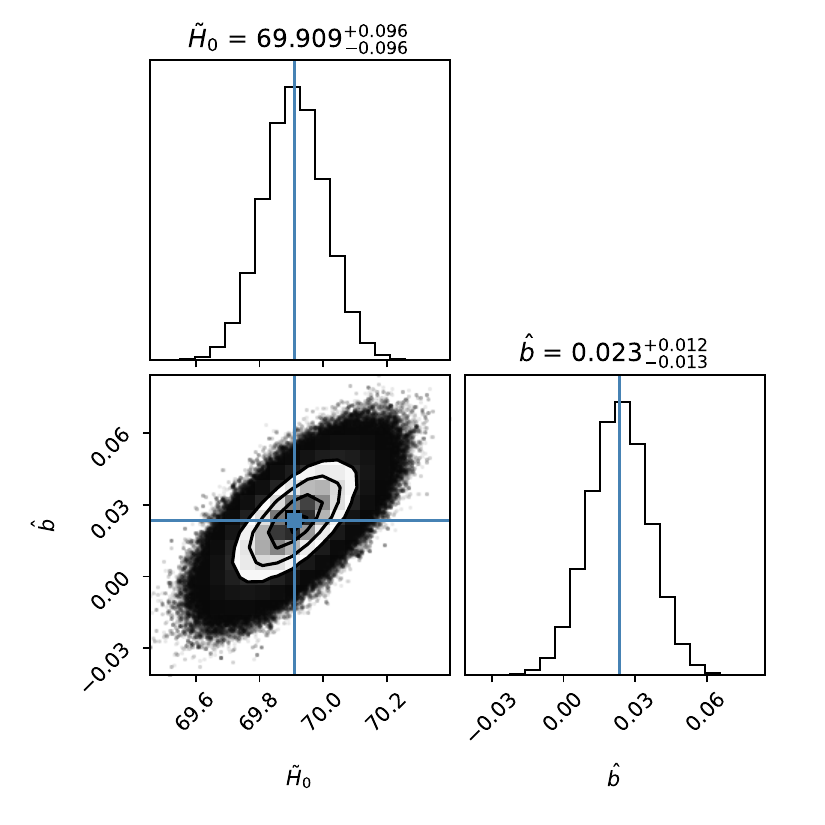}
        \vspace{0.15cm}

        (a)
    \end{minipage}\hfill
    \begin{minipage}{0.46\textwidth}
        \centering
        \includegraphics[width=\linewidth]{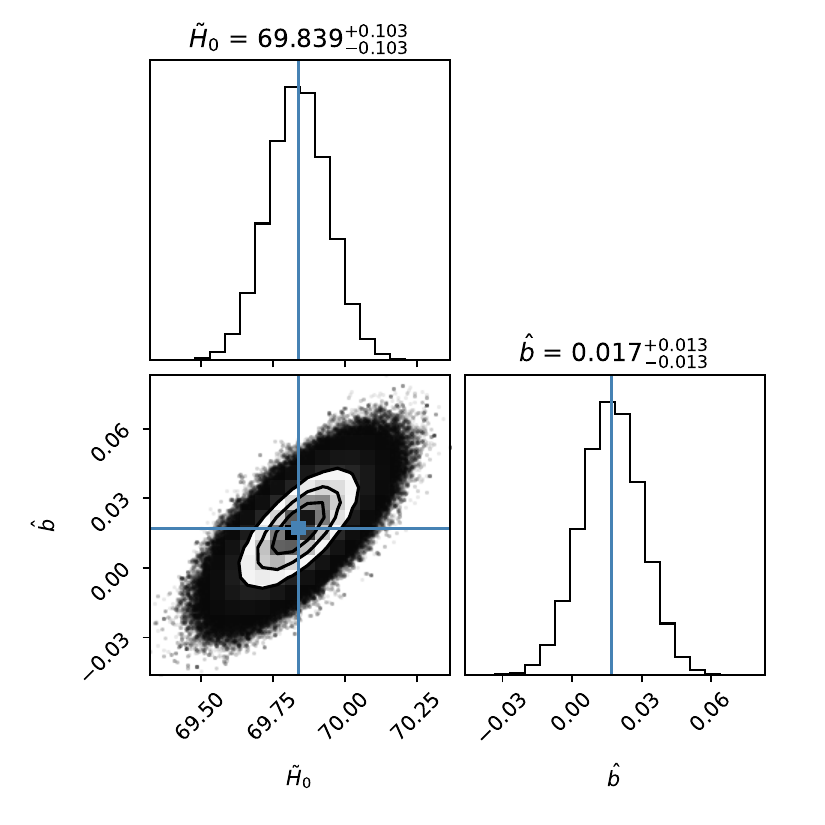}
        \vspace{0.15cm}

        (b)
    \end{minipage}

    \vspace{0.35cm}

    \begin{minipage}{0.46\textwidth}
        \centering
        \includegraphics[width=\linewidth]{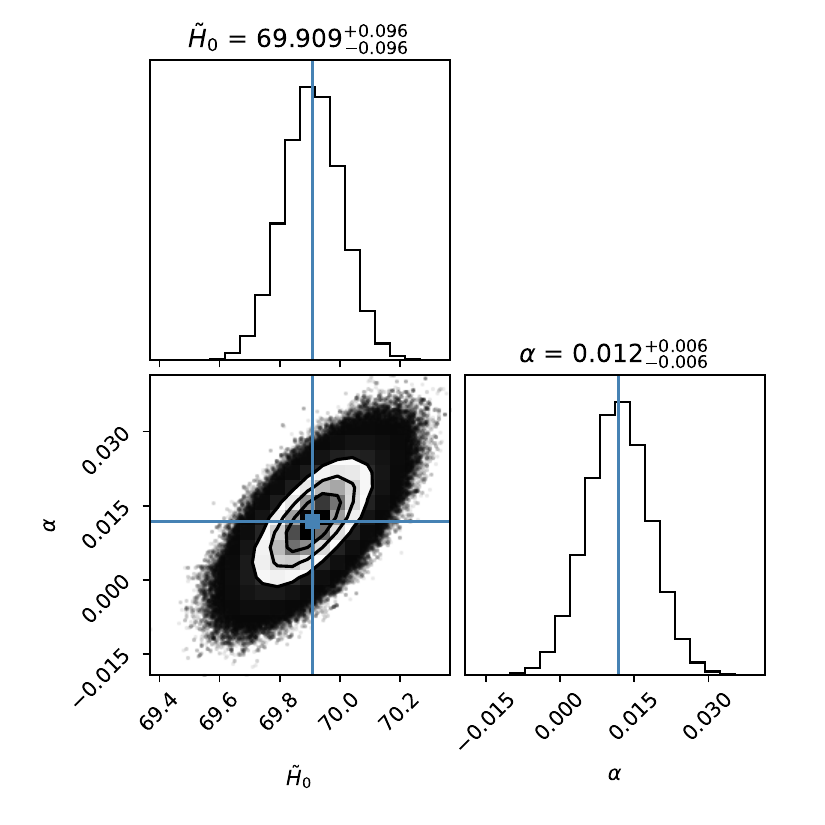}
        \vspace{0.15cm}

        (c)
    \end{minipage}\hfill
    \begin{minipage}{0.46\textwidth}
        \centering
        \includegraphics[width=\linewidth]{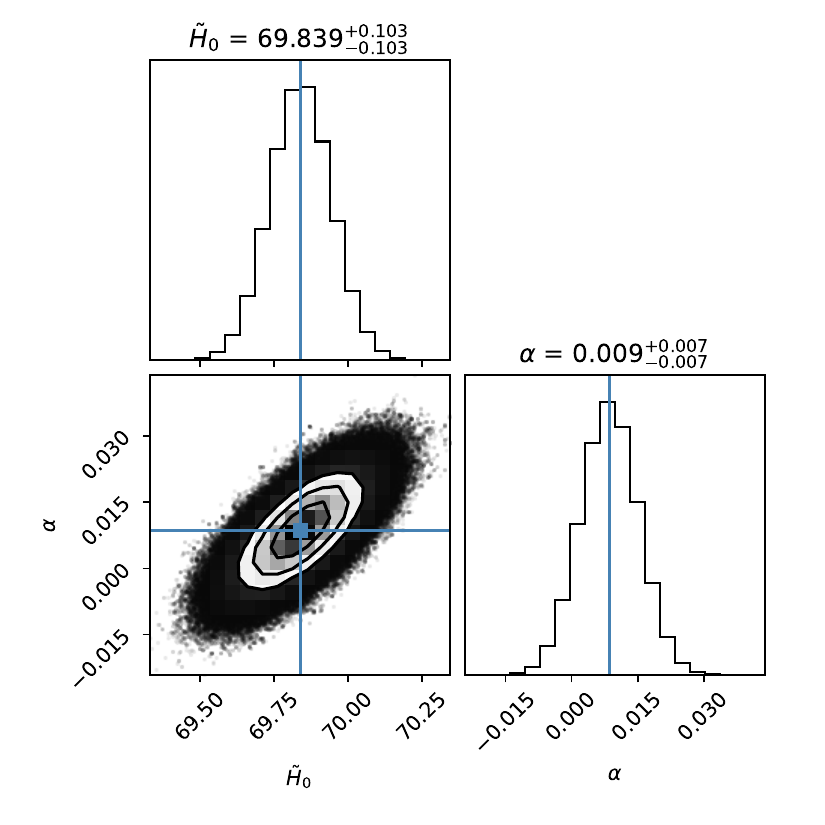}
        \vspace{0.15cm}

        (d)
    \end{minipage}

    \vspace{0.4cm}

    \caption{
    Corner plots showing the posterior distributions of the model parameters obtained from the MCMC analysis of the Master SNe~Ia Sample. 
    Figures (a) and (b) correspond to the logarithmic parameterization of the Hubble constant, with free parameters $\tilde{H}_0$ and $\hat{b}$, using datasets including and excluding low-redshift SNe~Ia, respectively. 
    Figures (c) and (d) show the corresponding results for the power-law parameterization, characterized by the parameters $\tilde{H}_0$ and $\alpha$. 
    The contours denote the $68\%$ and $95\%$ confidence regions, with the marginalized one-dimensional posterior distributions shown along the diagonal.
    }
    \label{fig:corner_plots}
\end{figure}

\begin{figure*}[p]
    \centering

    \includegraphics[width=0.72\textwidth]{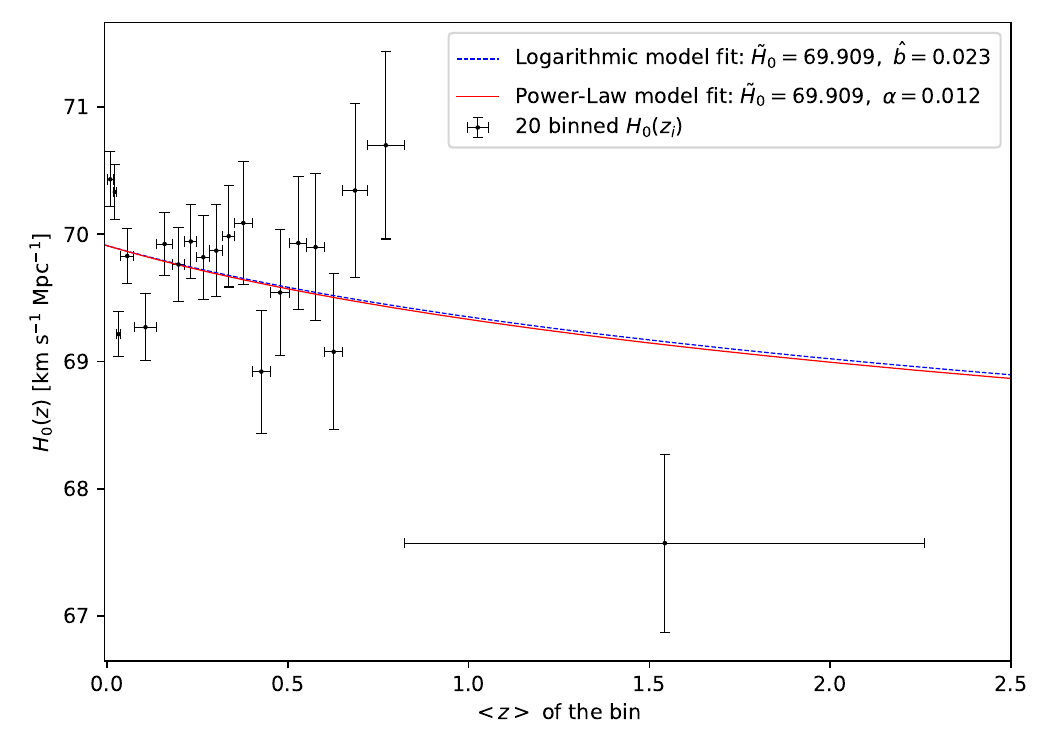}
    \vspace{0.15cm}
    \\
    (a)
    \vspace{0.2cm}

    \includegraphics[width=0.72\textwidth]{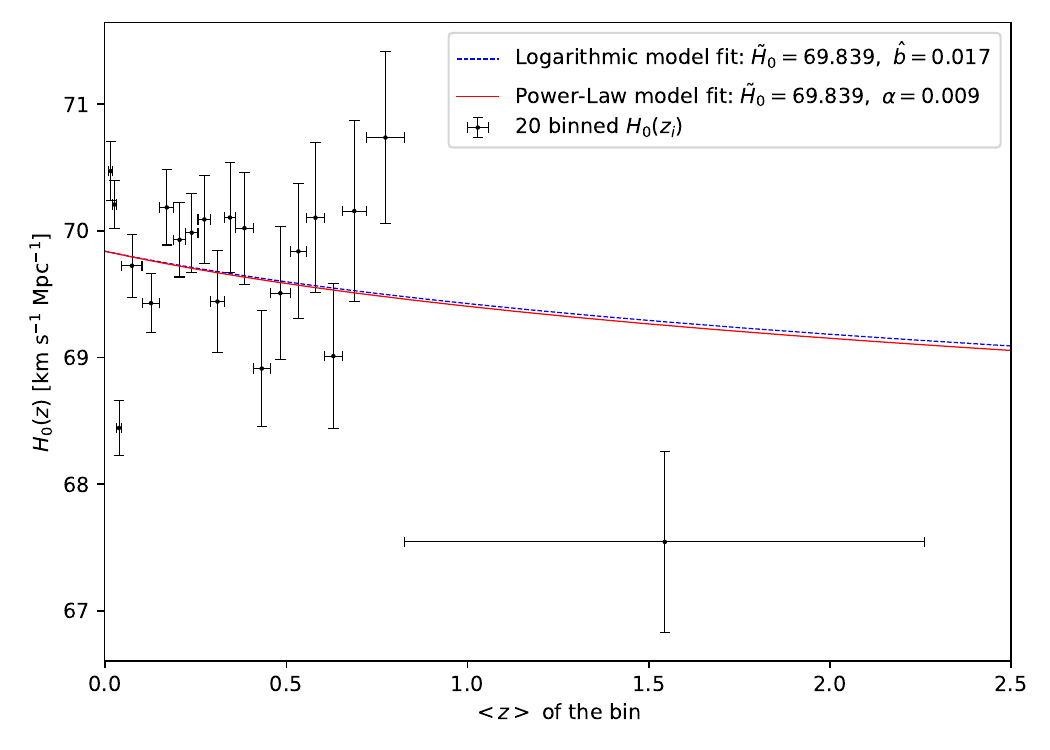}
    \vspace{0.15cm}
    \\
    (b)

    \caption{
    Best-fit model predictions for the Hubble constant reconstructed from the \textbf{20-bin analysis} as a function of redshift, obtained from the Master SNe~Ia Sample. Figures (a) and (b) correspond to the logarithmic parameterization, $H_{0,\log}$, and the power-law parameterization, $H_{0,\mathrm{pl}}$, using datasets including and excluding low-redshift SNe~Ia, respectively. The black points represent the binned estimates of $H_0({z_i})$, with associated uncertainties. The blue dashed line shows the best-fit prediction of the logarithmic model, while the red solid line shows the best-fit prediction of the power-law model. The quantity $\langle z \rangle$ denotes the mean redshift in each bin obtained from the 20-bin analysis.
    }
    \label{fig:best_fit_log-pl}
\end{figure*}

Using the 20-bin observational $H_{0}(z_i)$ measurements derived from the binned Master SNe Ia Sample, both including and excluding the low-$z$ SNe Ia, we constrain the logarithmic and power-law parameterizations given by Eqs.~\eqref{H0_log} and~\eqref{2.4Eq}. Specifically, we estimate the common normalization parameter $\tilde{H}_0$ together with the model parameter $\hat{b}$ for the logarithmic form or $\alpha$ for the power-law form by minimizing the $\chi^2$ statistic defined in Eq.~\eqref{chi2_mat} using MCMC sampling. The corner plots showing the posterior distributions of the model parameters inferred from the MCMC analysis are shown in \autoref{fig:corner_plots}. For both the logarithmic and power-law parameterizations, we determine the best-fit evolution of the effective Hubble constant as a function of redshift. The resulting $H_0(z_i)$ are shown in \autoref{fig:best_fit_log-pl} for the logarithmic and power-law parameterizations (shown in dashed and solid lines respectively), using datasets including and excluding low-redshift SNe~Ia, respectively. Moreover, we compute the reduced chi-squared, $\chi^2_{\mathrm{red}}$ \citep{bevington2003data}, and the Bayesian Information Criterion (BIC; \citet{2007MNRAS.377L..74L}) for both parametrizations, and data sets, to assess the goodness of fit and the relative statistical preference between these models.The reduced chi-squared is defined as $\chi^2_{\rm red}=\chi^2_{\rm min}/(N-k)$, and the Bayesian Information Criterion as $\mathrm{BIC}=\chi^2_{\rm min}+k\ln N$, where $\chi^2_{\rm min}$ is the minimum value of the $\chi^2$ statistic, $N=20$ is the number of redshift bins, and $k=2$ is the number of free parameters ($\tilde{H}_0$ and $\hat{b}$ or $\alpha$). This approach enables a direct comparison, summarized in \autoref{tab:combined_apj_final}, of the logarithmic and power-law forms of the Hubble constant across the explored redshift range.

\section{Discussion of our results}
\label{sec:Discussion}

Both samples (the Master sample with and without low-z SNe Ia) satisfy the validity requirements of the small-redshift approximation relevant for testing the equivalence condition, since the terms for both the logarithmic and power-law parameterizations, 
$\left|\alpha \ln(1+z)\right|$ and $\left|\hat{b} \ln(1+z)\right|$, 
are of the order $10^{-5}$ to $10^{-2}$ in the redshift regime considered.
Here, $H_0(z)$ is treated as an inferred, bin-derived quantity.
The best-fit values of the free parameters obtained from the two parameterizations are summarized in \autoref{tab:combined_apj_final}, where for the Master Sample including low-$z$ SNe~Ia the best-fit values are $\alpha = 0.012$, $2\alpha = 0.024$, and $\hat{b} = 0.023$, while excluding low-$z$ SNe~Ia, $\alpha = 0.009$, $2\alpha = 0.018$, and $\hat{b} = 0.017$. This agreement within $1\sigma$ indicates that the fitted parameters satisfy the approximate relation $\hat{b} \simeq 2\alpha$ in both cases. Both parameterizations yield compatible values within 1 $\sigma$ of $\tilde{H}_0$, with $69.909^{+0.096}_{-0.096}$ (with low-$z$ data) and $69.839^{+0.103}_{-0.103}$ (without low-$z$ data).
Consequently, the reconstructed Hubble constant satisfies 
$H_{0,\log}(z) \simeq H_{0,\mathrm{pl}}(z)$, 
indicating that, within the redshift range examined in this work, the logarithmic and power-law parameterizations are equivalent at leading order in $\ln(1+z)$.
Thus, the two parameterizations are statistically consistent within uncertainties
over the redshift range considered, with no significant deviation observed between them.
Extending this analysis to substantially higher redshifts will be essential for testing whether this equivalence persists beyond the regime probed here. 

The logarithmic parameterizations yield nearly identical BIC values for both datasets, with $\Delta\mathrm{BIC}=0.023$ for the analysis including the low-$z$ sample and $\Delta\mathrm{BIC}=0.015$ when the low-$z$ sample is excluded. This comparison is made among the two models without considering any of the two as a reference model. According to the standard interpretation of the Bayesian Information Criterion \citep{Kass01061995}, these differences provide no evidence favoring one parameterization over the other. In contrast, excluding the low-$z$ sample results in a substantial increase in both $\chi_{\mathrm{red}}^{2}$ and the BIC ($\Delta \mathrm{BIC}_{Log}=30.109$ for the logarithmic and $\Delta \mathrm{BIC}_{PL}=30.101$ for the power-law parameterization, respectively), indicating that the inclusion of the low-redshift supernovae leads to a significantly better overall description of the data. In \autoref{tab:combined_apj_final}, the value of $H_{0,\log}(z=10^{20})$ is not reported for the analysis including the low-$z$ sample because the best-fit logarithmic model predicts a maximum redshift, $z_{\max}\simeq3.3\times10^{18}$, corresponding to a minimum scale factor; therefore, the extrapolation is limited to $z\leq z_{\max}$. As discussed in \autoref{2.3}, Eq.~\eqref{amin} gives $z_{\max}=e^{1/\hat{b}}-1$, implying that $z_{\max}$ depends exponentially on the fitted parameter $\hat{b}$ and is therefore highly sensitive to even small variations in $\hat{b}$. For $\hat{b}=0.02$, $z_{\max}\approx5\times10^{21}$, corresponding to the inflationary redshift regime; however, a detailed discussion of the extrapolation in this regime is beyond the scope of the present work.

The value $\hat{b} \simeq 0.02$ for the logarithmic model was previously inferred by \citet{2025arXiv250902636L} using the single CMB data point at $z = 1100$ (see: \autoref{2.3}). Following the same approach, we estimate $\alpha \simeq 0.01$ for the power-law model (see: \autoref{2.4}). Our analysis using the Master sample SNe~Ia dataset is consistent with these values, indicating that the models remain viable at least up to $z \sim 1100$.
Since the logarithmic and power-law models are nearly indistinguishable over the redshift range examined, we extend our analysis by extrapolating the logarithmic and power-law parameterizations of the Hubble constant, $H_{0,\log}(z)$ and $H_{0,\mathrm{pl}}(z)$, to very high redshifts to investigate their behavior beyond the observationally accessible regime.
In particular, we explore their evolution up to $z \sim 1100$, corresponding to the epoch of cosmic microwave background (CMB) formation, then to $z \sim 10^{9}$, corresponding to the epoch of Big Bang nucleosynthesis (BBN), and further extend the extrapolation to the range $z \sim 10^{20}$, relevant to the inflationary era. This allows us to examine the asymptotic properties and consistency of the adopted parameterizations in the extreme high-redshift limit. All results from the high-redshift extrapolation of $H_{0,\log}(z)$ and $H_{0,\mathrm{pl}}(z)$, corresponding to the logarithmic and power-law parameterizations, respectively, are presented in \autoref{tab:combined_apj_final}. 
We note that the uncertainties increase with redshift, particularly in the extrapolated regime. These extrapolations should be regarded as qualitative probes of theoretical consistency rather than direct observational constraints, as they extend significantly beyond the redshift range accessible to current data.

\begin{table*}[t]
\centering
\small
\renewcommand{\arraystretch}{1.2}

\begin{tabular}{llcc}
\hline
\textbf{Model} & \textbf{Parameter} & \textbf{With Low-$z$} & \textbf{Without Low-$z$} \\
\hline


Logarithmic & $\tilde{H}_0$ 
& $69.909^{+0.096}_{-0.096}$ & $69.839^{+0.103}_{-0.103}$ \\

& $\hat{b}$ 
& $0.023^{+0.012}_{-0.013}$ & $0.017^{+0.013}_{-0.013}$ \\

& $\chi^2_{\mathrm{red}}$ 
& $1.242$ & $2.079$ \\

& BIC 
& $50.712$ & $80.821$ \\
& $\Delta\mathrm{BIC}$ & $0.023$ & $0.015$\\

& $H_{0,\log}(z=1100)$ 
& $63.914^{+3.225}_{-3.387}$ & $65.536^{+3.223}_{-3.379}$ \\

& $H_{0,\log}(z=10^{9})$ 
& $50.132^{+11.154}_{-13.798}$ & $56.137^{+10.521}_{-12.779}$ \\

& $H_{0,\log}(z=10^{20})$ 
& \textemdash & $32.231^{+11.774}_{-14.528}$ \\

& $\ln(z_{\max})$ 
& $42.650^{+1.345}_{-0.330}$ & $58.517^{+3.272}_{-0.430}$\\

\hline


Power-law & $\tilde{H}_0$ 
& $69.909^{+0.096}_{-0.096}$ & $69.839^{+0.103}_{-0.103}$ \\

& $\alpha$ 
& $0.012^{+0.006}_{-0.006}$ & $0.009^{+0.007}_{-0.007}$ \\

& $\chi^2_{\mathrm{red}}$ 
& $1.243$ & $2.079$ \\

& BIC 
& $50.735$ & $80.836$ \\
& $\Delta\mathrm{BIC}$ & $0.023$ & $0.015$\\

& $H_{0,\mathrm{pl}}(z=1100)$ 
& $64.393^{+2.861}_{-2.748}$ & $65.778^{+2.986}_{-2.866}$ \\

& $H_{0,\mathrm{pl}}(z=10^{9})$ 
& $54.807^{+7.633}_{-6.719}$ & $58.473^{+8.341}_{-7.321}$ \\

& $H_{0,\mathrm{pl}}(z=10^{20})$ 
& $40.696^{+13.727}_{-10.284}$ & $47.054^{+16.301}_{-12.134}$ \\

\hline
\textbf{Dataset comparison} & $\Delta\mathrm{BIC}_{Log}$ & $30.109$  & \textemdash\\
& $\Delta\mathrm{BIC}_{PL}$ & $30.101$ & \textemdash \\
\hline
\end{tabular}

\caption{
Best-fit parameters from the 20-bin analysis of the Master SNe~Ia Sample and high-redshift extrapolation of $H_0(z)$ for logarithmic and power-law models. Results are shown for analyses performed with and without the low-redshift subsample. The last rows show the $\Delta\mathrm{BIC}$ within the same model but among the two different data sets.}
\label{tab:combined_apj_final}
\end{table*}


\clearpage


\section{Discussion in relation to the literature}
\label{literature}
The $H_0(z)$ has also been investigated by other authors in the literature in different ways. \cite{2025arXiv251014390W} perform a model-independent, non-parametric reconstruction of the dark-energy equation of state together with an evolving Hubble constant using DESI DR2 BAO and Type Ia supernova data. Their analysis finds clear redshift evolution in both $w(z)$ and $H_0(z)$, with $H_0$ decreasing from low to high redshift, a behavior consistent with the trend obtained in our redshift-binned reconstruction of the effective $H_0$. They further show that the joint $w(z)$–$H_0(z)$ reconstruction is statistically favored over the standard $\Lambda$CDM framework, indicating that such evolution may help alleviate the Hubble constant tension.
Complementary analyses by \cite{Krishnan2020PhRvD.102j3525K}, employing Gaussian process reconstructions, provided a model-independent perspective, suggesting that deviations from strict $\Lambda$CDM expectations may arise when late-time expansion data are treated non-parametrically.
\cite{Kazantzidis2020PhRvD.102b3520K,Alestas2020PhRvD.101l3516A} found a variation of the absolute Magnitude, the $M$ parameter, as a function of the redshift and since $M$ and $H_0$ are degenerate, this is equivalent to our finding of a varying $H_0$. More recently, \cite{Xu2024MNRAS5305091} examined combined SNe Ia and $H(z)$ datasets, again finding that mild redshift-dependent behavior cannot be trivially excluded. \cite{2025JHEAp..45..290D} broadened the investigation of a redshift-dependent Hubble constant by introducing a comparison between the power-law trend and the Jesus parametrization \citep{Jesus2018arXiv1712.01075}, i.e., a specific phenomenological prescription for the redshift evolution of the Hubble constant. Within this framework, the parametrization provides an alternative description of an effective $H_0(z)$, allowing direct comparison with other commonly adopted functional forms. Their analysis showed that, despite its theoretical motivation, the Jesus parametrization is statistically disfavoured relative to the power-law parameterization. 
\cite{2025ApJ...994L..22J} analyze non parametrically the DESI baryon acoustic oscillation measurements combined with Type Ia supernovae and find that the derived dark energy equation of state evolves with redshift, which in turn yields an effective $H_0(z)$ that decreases with redshift and can alleviate the Hubble tension within a unified dynamical framework, thus confirming our findings. Interestingly, \cite{2025arXiv251104610E} found a trend in $H_0$ as a function of redshift that is also recovered in other domains of astrophysics, such as by using Fast Radio Bursts, \citep{FastRadioBursts2026ApJ...996...50K}, showing an indication of a deeper inadequacy in the $\Lambda$CDM model.
However, there are caveats also discussed in the literature about this trend. Indeed, \cite{Mo2026arXiv260115765M} performed a joint redshift-binned analysis of $H_0$ using multiple late-time probes and found that the binned variations are consistent with parameter degeneracies, possibly indicating no statistically significant evidence for an intrinsic redshift evolution of the Hubble constant. Continuing on these caveats, \cite{2025NewA..12102454S} conducted an analysis of observational data under different cosmological models and reported that a mild redshift dependence of the Hubble constant can arise when fitting cosmological parameters in narrow redshift slices. They find that certain parameter combinations — including $H_0$  — show systematic trends with redshift that may reflect model degeneracies, dataset selection effects, or hints of physics beyond a simple $\Lambda$CDM description.
Similarly, \cite{Liu2025PhRvD112L3539} find model-independent $H_0$ estimates at several redshifts that show no statistically significant evolution, offering an independent geometric constraint on $H_0$.

On the other hand, \citet{2025MNRAS.542.1063H} constrained the transition redshift---the epoch at which the cosmic expansion changed from deceleration to acceleration---using the latest $H(z)$ measurements, showing that while the reconstructed expansion history remains broadly consistent with late-time acceleration, the results highlight the sensitivity of cosmological parameter inferences, including implications for the Hubble tension, to dataset selection and reconstruction methodology. 

Some authors, such as \cite{2025arXiv251212697Z}, consider that a parameterized dark energy equation of state which allows additional dynamical freedom in the late-time expansion can help reconcile Hubble parameter measurements across probes.

 \citet{2025JHEAp..4700374M} investigated cosmic evolution within the framework of $f(R,T)$ modified gravity, showing that deviations from standard General Relativity can naturally generate effective dark energy behavior and alter the late-time expansion dynamics. Actually,
in this study the shape
of the Universe matter source is altered, leading
to a modified form of the corresponding equation of state. The theoretical predictions are then compared with SN Ia and BAO data to constrain the
free model parameters. Also, from this
analysis one could recover an effective Hubble constant, as
a measure of the emerging discrepancy
with respect to the standard
$\Lambda$CDM-model.

In relation to the combination of components, \citet{2025PhRvD.112f3517Y} studied a cosmological scenario combining an early dark energy component with an interacting dark sector model, finding that while the joint framework can alleviate both the $\sigma_8$ and $H_0$ tensions, it does not provide a complete resolution, highlighting the persistent challenges in simultaneously reconciling late-time expansion and structure growth observables. Interestingly, \citet{2025arXiv250709228L} investigated Torsion Condensation (TorC), an extension of gravity based on Poincar\'e gauge theory incorporating intrinsic torsion degrees of freedom, and found that the model permits a higher inferred value of $H_0$, thereby alleviating the tension between Planck and SH0ES measurements, although the improvement is not sufficient to decisively favour TorC over $\Lambda$CDM in Bayesian model comparison. \citet{2026arXiv260208928A} present a data-driven reconstruction of the late-time expansion history indicating a sign-changing effective dark-energy density and possible phantom/quintom behavior, while remaining overall consistent with flat $\Lambda$CDM in Bayesian evidence. Notably, \citet{2026arXiv260105218S} propose a QFT-based changing vacuum model in which dynamical vacuum energy drives inflation without an inflaton field, unifies early- and late-time cosmic dynamics, and may help alleviate the Hubble tension.

In future studies, examining the logarithmic parametrization within broader cosmological frameworks and performing a systematic comparison with the power-law parametrization may provide a more comprehensive assessment of their respective physical implications.

\section{Summary and Conclusions}
\label{sec:Conclusions}
This work provides a direct empirical assessment of logarithmic and power-law parameterizations of the Hubble constant using the binned Master Sample, identifying the redshift interval over which the two models yield equivalent predictions and quantifying the regime where deviations emerge.
For each case, we determine the best-fitting cosmological parameters, specifically the Hubble constant $H_0$ and the present matter density parameter $\Omega_{m0}$. Using the resulting binned constraints, we then examine the redshift dependence of the Hubble constant through both a power-law and a logarithmic parameterization. Within the redshift range probed by the current data, the two parameterizations are found to be equivalent under the condition, $\hat{b}=2\alpha$.

To further investigate their behaviour beyond the observed regime, we extrapolate both parameterizations to three characteristic epochs of the early Universe: $z=1100$ (CMB decoupling), $z=10^{9}$ (BBN), and $z=10^{20}$ (inflationary era). It is evident that
the power-law and logarithmic parametrizations of $H_{0,\log}(z)$ and $H_{0,\mathrm{pl}}(z)$ remain statistically consistent reconstructions over a broad redshift interval. 
When extrapolated to the CMB scale, the inferred values 
of $H_{0,\log}(z)$ and $H_{0,\mathrm{pl}}(z)$ differ by only 
$\sim 0.75\%$ (with low-$z$ data) and $\sim 0.37\%$ (without low-$z$ data) using the Master Sample. 
At the epoch of BBN these differences increase to $\sim 9.32\%$ 
with low-$z$ data and $\sim 4.17\%$ without low-$z$ data, and further grow to $\sim 46\%$ at 
ultra-high redshifts, reaching the inflationary era when using the Master Sample 
without low-$z$ data. 

For the logarithmic parametrization, $H_{0,\log}(z)$,   as explained in \autoref{2.3},   there exists a maximal redshift  $\zmax$,  corresponding to 
a minimal scale factor $\amin$,  and the solution to $a(t)$ extends to before such a time $\tmin$ such that a Big Bang singularity associated with $a(t) =0$ is avoided. 
Based on our analysis,   
$\ln(z_{\max})  \approx 42.650^{+1.345}_{-0.330}$ (with low-$z$ data) and $\ln(z_{\max})  \approx 58.517^{+3.272}_{-0.430}$ 
(without low-$z$ data). While $H_{0,\mathrm{pl}}(z)$ remains well defined under 
formal extrapolation in all three redshift regimes,   $H_{0,\log}(z)$ exhibits a vanishing at a 
finite redshift, whereas $H_{0,\mathrm{pl}}(z)$ approaches zero asymptotically 
as, $z \rightarrow \infty$. This distinction implies that the logarithmic parametrization 
admits a potential resolution of the Big Bang singularity, while the power-law parametrization is in agreement with the standard theory, which includes the existence of Big Bang singularities.

Here we also investigate the impact of peculiar velocities on distance measurements in the nearby Universe \citep{2006PhRvD..73l3526H,2011ApJ...741...67D,2022ApJ...938..112P}, as discussed in \autoref{sec:Data}. This analysis is particularly relevant because we employ the binned Type~Ia supernovae Master Sample \citep{2025JHEAp..4800405D}, performing the cosmological fits both including and excluding the low-redshift data. From \autoref{tab:combined_apj_final}, it is visible that including the low-redshift supernovae in the binned Master Sample leads to a slightly better fit quality for both the logarithmic and power-law models. 
The high-redshift extrapolation of both the logarithmic and power-law parameterizations of the effective Hubble constant exhibits a clear sensitivity to the inclusion of low-$z$ data. When low-redshift supernovae are excluded, both $H_{0,\log}(z)$ and $H_{0,\mathrm{pl}}(z)$ yield systematically larger values at increasing redshift. For the logarithmic model the increase is $2.53\%$ at $z=1100$ and $11.97\%$ at $z=10^{9}$, while for the power-law model the increase is $2.14\%$ at $z=1100$, $6.70\%$ at $z=10^{9}$, and $15.63\%$ at $z=10^{20}$.
The quoted percentage differences expressed with $\Delta\mathrm{BIC}$ are based on the best-fit values, and are intended to quantify statistically the relative difference between the datasets with and without the low-$z$ sample and the two models, while the associated uncertainties are presented in \autoref{tab:combined_apj_final}.

These results provide a quantitative characterization of the redshift regime over which phenomenological parameterizations of $H_0(z)$ remain observationally consistent and degenerate, while identifying the redshift scales where model discrimination becomes possible.
One of the key findings of this work is that, within the redshift range explored in our analysis, the logarithmic and power-law parameterizations of the redshift-dependent Hubble constant are effectively equivalent under the condition $\hat b = 2\alpha$, despite their different theoretical motivations. 

At higher redshift, the two parameterizations exhibit distinct extrapolation behaviour. At the same time, both approaches may provide an indicative improvement over current studies of the Hubble constant tension. 
More broadly, this analysis highlights how phenomenological equivalence at low redshift does not guarantee similar cosmological extrapolation behavior to the very high-z regime.

In light of these findings, future studies may incorporate independent high-redshift observables, such as GRBs \citep{2023MNRAS.518.2201D,2022PASJ...74.1095D,2015MNRAS.451.3898D,2017A&A...600A..98D,2024JHEAp..44..323F,2026JHEAp..4900439M}, Quasars \citep{2023ApJ...951...63D,2023ApJ...950...45D}, Cosmic Chronometers \citep{2010JCAP...02..008S,2012JCAP...08..006M,2018JCAP...04..051G,2024PhLB..85839027F} at intermediate and high redshifts, BAOs measured from the Ly$\alpha$ forest \citep{2013A&A...552A..96B,2013JCAP...04..026S}, and strong gravitational lensing time-delay systems \citep{2019MNRAS.484.4726B,2020ApJ...895L..29L,2020A&A...643A.165B}. These probes provide complementary access to the cosmic expansion history at $z \gtrsim 2$ and beyond, thereby offering additional leverage to assess potential deviations between $H_{0,\log}(z)$ and $H_{0,\mathrm{pl}}(z)$ at earlier epochs. Furthermore, exploring these parameterizations within alternative cosmological models beyond the standard $\Lambda$CDM framework — including extensions such as $f(R)$ modified gravity — would allow a broader assessment of their behavior under different cosmological assumptions and help clarify whether the quantitative distinctions observed in their high-redshift extrapolations carry physical implications.

\appendix
\numberwithin{equation}{section}

\section{Statistical $\chi^2$ Evaluation and Matrix Formalism} 
\label{sec:AppendixA}

To evaluate the goodness-of-fit for the models discussed, we employ the $\chi^2$ statistic. Consider a general model function $y(x)$ expressed as an expansion over a set of basis functions $f_k(x)$:
\begin{equation}
y(x) = \sum_{k} a_k f_k(x)
\end{equation}
where, the variable $x$ denotes the independent quantity of the data, and $a_k$ represents the coefficient parameter corresponding to the function $f_k(x)$. For a dataset consisting of $n$ independent observations $(x_i, y_i)$, where $i = 1, 2, \ldots, n$, and each $y_i$ has an associated measurement uncertainty $\sigma_i$, the standard definition of the multidimensional chi-square (see: \citet{bevington2003data,Press:2007ipz}) statistic is given by:

\begin{equation}
\begin{aligned}
\label{chi_general}
\chi^2 &= \sum_{i=1}^{n} \left[ \frac{y_i - y(x_i)}{\sigma_i} \right]^2 \\
&= \sum_{i=1}^{n} \left[ \frac{1}{\sigma_i} \left( y_i - \sum_{k} a_k f_k(x_i) \right) \right]^2
\end{aligned}
\end{equation}
This expression can be equivalently derived using matrix formalism.

We define the residual vector, $R$ as the difference between the obtained data vector $y$ and the model prediction vector $y(x)$:
$R = y - y(x)$.
The components of the residual vector are, $R_i = y_i - y(x_i)$.\\
We introduce the covariance matrix 
$C$, defined for uncorrelated measurements \citep{bevington2003data,Press:2007ipz} as:
\begin{equation}
\label{covariance}
C = \mathrm{diag}\,(\sigma_1^2,\, \sigma_2^2,\, \dots,\, \sigma_n^2).
\end{equation}

Let $M$ be an $n \times n$ symmetric matrix with elements $M_{ij}$.\\
The quadratic form (See: \citet{strang2006linear,horn2012matrix}) $R^T M R$ is expressed as,
\begin{equation}
    R^T M R 
    = \sum_{i=1}^{n} \sum_{j=1}^{n} R_i\, M_{ij}\, R_j .
\end{equation}
Taking $M = C^{-1}$, we may write,
\begin{align}
\begin{aligned}
&[y - y(x)]^{T}\,C^{-1}\,[y - y(x)] \\
&\qquad = \sum_{i=1}^{n} \sum_{j=1}^{n} [y_i - y(x_i)]\, (C^{-1})_{ij}\, [y_j - y(x_j)] \\
&\qquad = \sum_{i=1}^{n} \sum_{j=1}^{n} [y_i - y(x_i)]
\left[ \delta_{ij}\,\frac{1}{\sigma_i^2} \right]
[y_j - y(x_j)] \\
&\qquad = \sum_{i=1}^{n} \frac{[\,y_i - y(x_i)\,]^2}{\sigma_i^2}\\
&\qquad = \sum_{i=1}^{n} \left[ \frac{1}{\sigma_i} \left( y_i - \sum_{k} a_k f_k(x_i) \right) \right]^2 \\
&\qquad = \chi^2
\end{aligned} 
\end{align}\\

For the logarithmic parameterization \citep{2025arXiv250902636L}, the Hubble constant is described by the expansion given in Eq.~\eqref{Log_expansion},
\begin{equation*}
\begin{aligned}
H_{0,\log}(z)
&=
\tilde{H}_{0}
\big[1-\hat{b}\ln(1+z)\big]^{1/2}
\\
&\simeq
\tilde{H}_{0}
\Big[
1
- \frac{\hat{b}}{2}\ln(1+z)
- \frac{\hat{b}^{2}}{8}\big(\ln(1+z)\big)^{2}
- \frac{\hat{b}^{3}}{16}\big(\ln(1+z)\big)^{3}
+ \cdots
\Big]
\end{aligned}
\end{equation*}
In the power-law parameterization \citep{2021ApJ...912..150D,2022Galax..10...24D,2025JHEAp..4800405D}, the Hubble constant is expressed through the expansion in Eq.~\eqref{PL_expansion},
\begin{equation*}
\begin{aligned}
H_{0,\mathrm{pl}}(z)
&=
\tilde{H}_{0}(1+z)^{-\alpha}
\\
&\simeq
\tilde{H}_{0}
\Big[
1
- \alpha\ln(1+z)
+ \frac{\alpha^{2}}{2}\big(\ln(1+z)\big)^{2}
- \frac{\alpha^{3}}{6}\big(\ln(1+z)\big)^{3}
+ \cdots
\Big]
\end{aligned}
\end{equation*}
In the logarithmic expansion, the Hubble constant is written as, 
\begin{equation}
\label{A.6}
H_{0,\log}(z)
= \tilde{H}_{0}
\big[1-\hat{b}\ln(1+z)\big]^{1/2} =
\sum_{k=1}^{\infty}
\hat{b}_k\, f_k(z) .
\end{equation}
where the expansion coefficients are $\hat{b}_1 = \tilde{H_0}$, $\hat{b}_2 = -\frac{\hat{b}}{2}\tilde{H_0}$, and so on.
Similarly, for the power-law expansion:
\begin{equation}
\label{A.7}
H_{0,\mathrm{pl}}(z) = \tilde{H}_{0}(1+z)^{-\alpha}= \sum_{k=1}^{\infty} \alpha_k\, f_k(z),
\end{equation}
with coefficients $\alpha_1 = \tilde{H_0}$, $\alpha_2 = -\alpha\tilde{H_0}$, and so on.

The basis functions in both parametrizations correspond to the powers of $\ln(1+z)$ in the Taylor series expansion with the general form being,
\begin{equation}
f_k(z) = (\ln(1+z))^{k-1}.
\end{equation}

For the 20-binned analysis (n=20), the obtained data vector is defined as,
$
y = H_0({\mathrm{z_i}}).
$
In the case of the logarithmic and power-law parameterizations of the Hubble constant, 
we consider the corresponding model prediction vector, 
as,
\begin{equation}
y(z) =
\begin{cases}
H_{0,\log}\!\left(z;\tilde{H}_0,\hat{b}\right),
& \text{for the logarithmic parameterization}, \\[6pt]
H_{0,\mathrm{pl}}\!\left(z;\tilde{H}_0,\alpha\right),
& \text{for the power-law parameterization}.
\end{cases}
\end{equation}

We define the model-specific residual vectors, $\Delta H_0$ as the difference between the obtained data and the theoretical model predictions:\\
For both models, the residual vector takes the form,
\begin{equation}
\Delta H_0^{\mathrm{model}}\!\left(\tilde{H}_0, \theta \right)
= H_0({z_i})- {H}_{0,model}\!\left(z;\tilde{H}_0, \theta\right).
\end{equation}
where $\theta=(\alpha, \hat{b})$ for the power law and logarithm parametrization, respectively.
The chi-square expression corresponding to the model can be written as,
\begin{align}
\chi^2_{model}
\newline
&=\left[
H_0({\mathrm{z_i}})
-
H_{0,model}\!\left(z;\tilde{H}_0,\theta\right)
\right]^{T}
C_{H_0}^{-1}
\left[
H_0({\mathrm{z_i}})
-
H_{0,}\!\left(z;\tilde{H}_0,\theta\right)
\right]
\notag \\[6pt]
&=
\left[
\Delta H_0^{model}\!\left(\tilde{H}_0,\theta\right)
\right]^{T}
C_{H_0}^{-1}
\left[
\Delta H_0^{model}\!\left(\tilde{H}_0,\theta\right)
\right]
\notag \\[6pt]
&=
\sum_{i=1}^{20}
\left[
\frac{
H_0({\mathrm{z_i}})
-
\sum_{k=1}^{\infty}
\theta_k\, f_k(z_i)
}{
\sigma_{H_0(z_i)}
}
\right]^2 
\notag \\[6pt]
&=
\sum_{i=1}^{20}
\left[
\frac{
H_0({\mathrm{z_i}}) - \tilde{H}_0\, g(z_i; \theta)
}{
\sigma_{H_0(z_i)}
}
\right]^2
\end{align}
where, $\sigma_{H_0(z_i)}$ denotes the uncertainty associated with the obtained Hubble constant, $H_0({z_i})$ in the $i$-th bin analysis, with redshift centered at $z_i$, for the 20-bin case with $i = 1, 2, \ldots, 20$. The covariance matrix ${C_{H_0}}$ is defined in Eq.~\eqref{Cmatrix}, and $H_{0,model}$ has been defined in Eqs.~\eqref{A.6} and \eqref{A.7}.

\section{Estimation of $\tilde{H}_0$ from $\chi^2$ Minimization} 
\label{sec:AppendixB}


Using the chi-square statistic in Eq.~\eqref{chi_general}, for the general parametrized model of the Hubble constant given in Eq.~\eqref{H0_model_general}, we obtain:
\begin{equation}
\chi^2_{\mathrm{model}} =
\sum_{i=1}^{n}
\left[
\frac{
H_0({\mathrm{z_i}}) - \tilde{H}_0\, g(z_i; \theta)
}{
\sigma_{H_0(z_i)}
}
\right]^2 \, .
\end{equation}

The minimum of $\chi^2_{\mathrm{model}}$ is obtained by requiring that its partial
derivatives with respect to each free parameter vanish \citep{bevington2003data}. In particular, minimizing
with respect to $\tilde{H}_0$ yields,
\begin{equation}
\label{eq.B1}
\frac{\partial \chi^2_{\mathrm{model}}}{\partial \tilde{H}_0} = 0 .
\end{equation}
The minimization condition of Eq.~\eqref{eq.B1}, then reduces to,
\begin{equation*}
-2
\sum_{i=1}^{n}
\frac{
g(z_i; \theta)
}{
\sigma_{H_0(z_i)}^2
}
\left[
{H}_0(z_i)
-
\tilde{H}_0
\, g(z_i; \theta)
\right]
=0 .
\end{equation*}
Solving for $\tilde{H}_0$, one obtains,
\begin{equation}
\label{eq.B4}
\tilde{H}_0
=
\frac{
\displaystyle
\sum_{i=1}^{n}
\frac{
g(z_i;\theta)\,
{H}_0(z_i)
}{
\sigma_{H_0(z_i)}^2
}
}{
\displaystyle
\sum_{i=1}^{n}
\frac{
g^2(z_i;\theta)
}{
\sigma_{H_0(z_i)}^2
}
} \, .
\end{equation}
For models satisfying the condition $g(z;\theta)=1$ at $z=0$, the normalization $\tilde{H}_0$ corresponds to the same value at the present epoch. Since, in our case, the function in the logarithmic model, $g_{\mathrm{log}}(z_i;\hat{b})$ in Eq.~\eqref{Eq.34}, and that in the power-law model, $g_{\mathrm{pl}}(z_i;\alpha)$ in Eq.~\eqref{Eq.39}, both equal unity at $z=0$, it follows that the normalization $\tilde{H}_0$ at $z=0$ is identical for these models and is consistent with $H_0$ within uncertainties.
\clearpage
\section{Joint Posterior Contour Plots for the 20 Redshift-Bin Analysis}
\label{sec:contours}
\begin{figure}[!ht]
    \centering

    \includegraphics[width=0.22\textwidth]{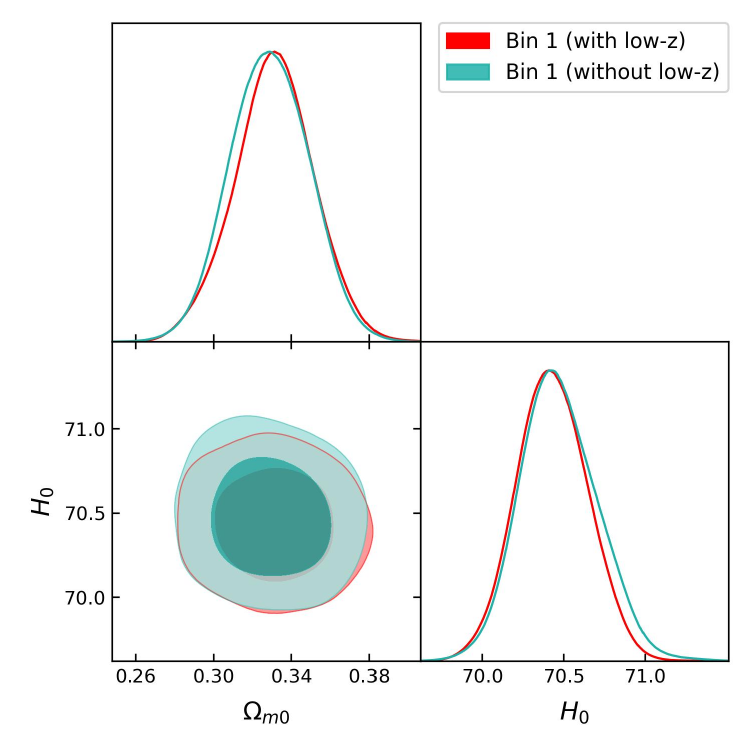}\hfill
    \includegraphics[width=0.22\textwidth]{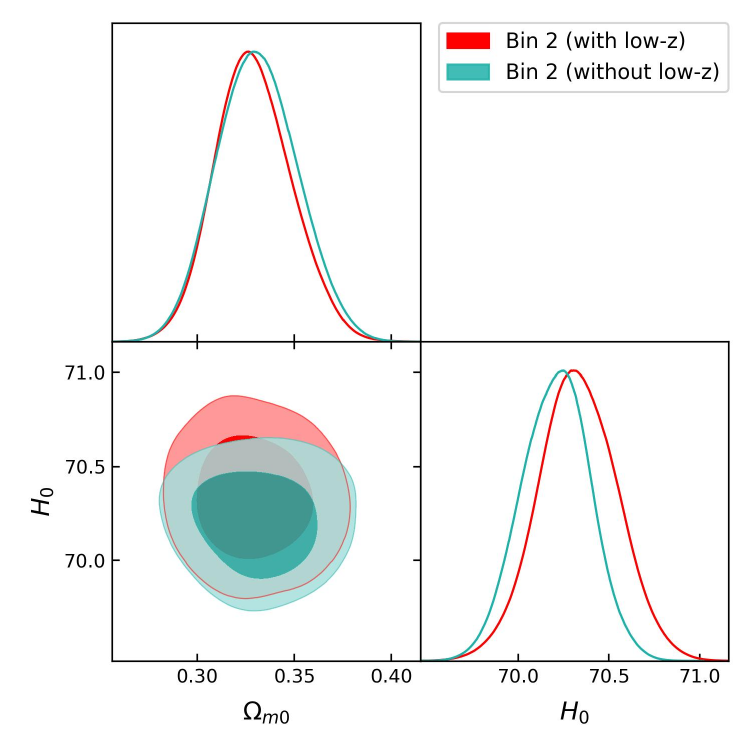}\hfill
    \includegraphics[width=0.22\textwidth]{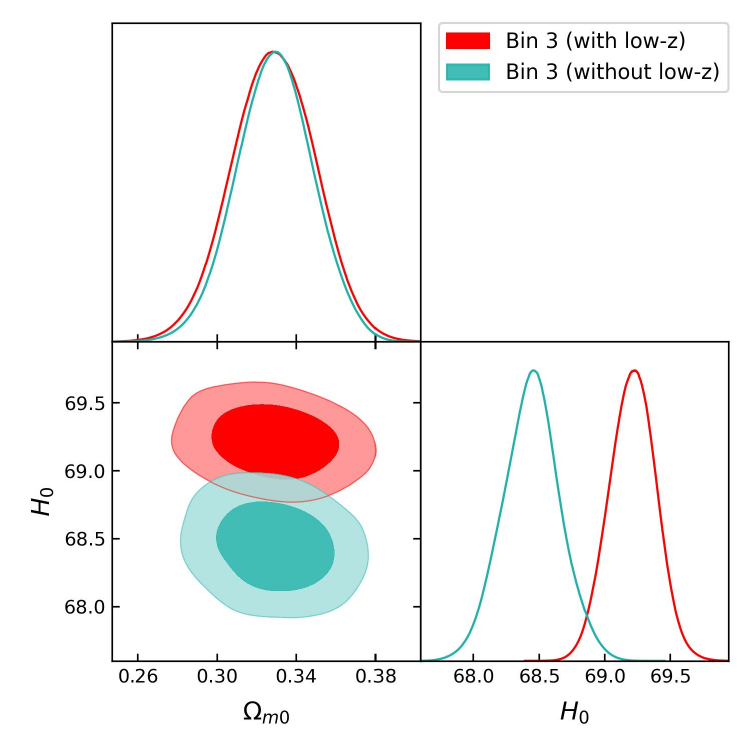}\hfill
    \includegraphics[width=0.22\textwidth]{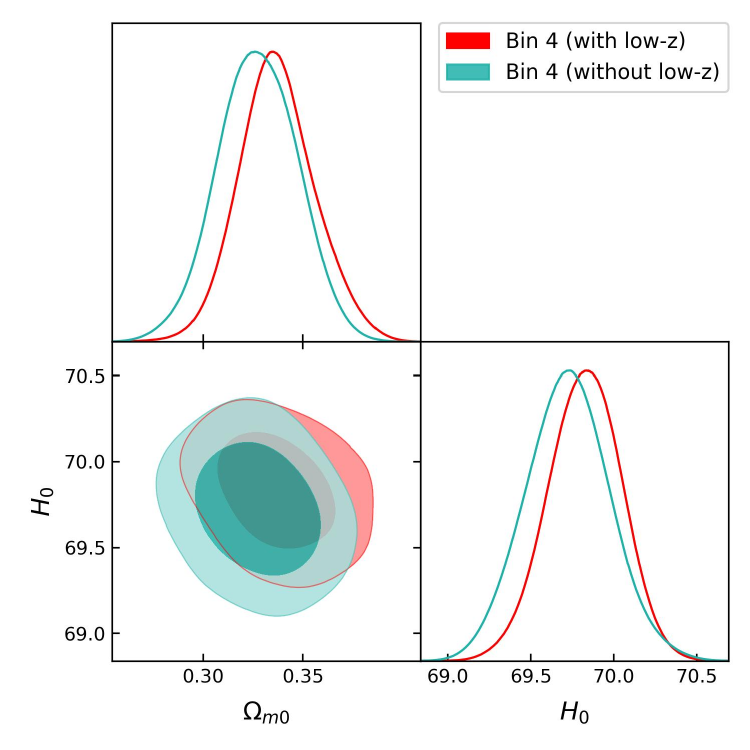}


    \includegraphics[width=0.22\textwidth]{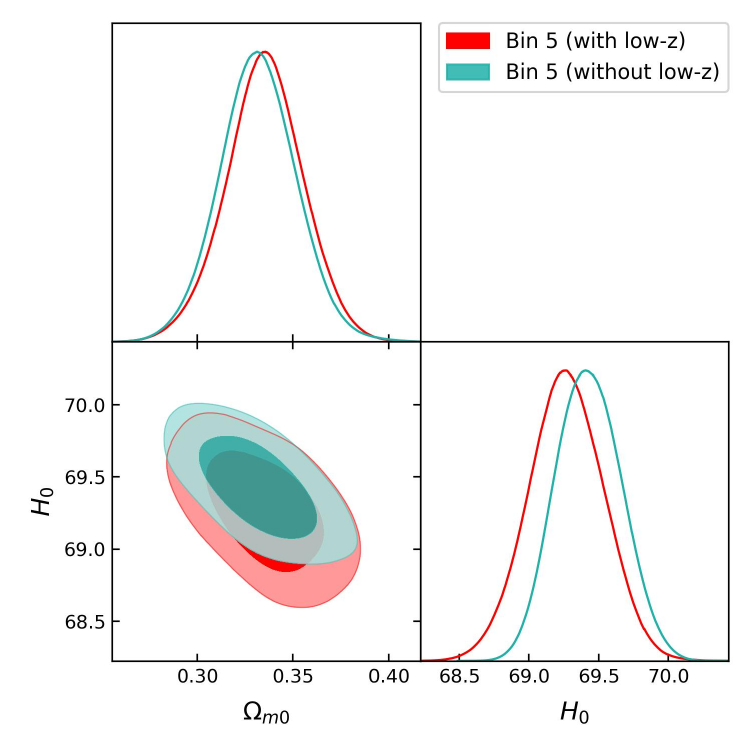}\hfill
    \includegraphics[width=0.22\textwidth]{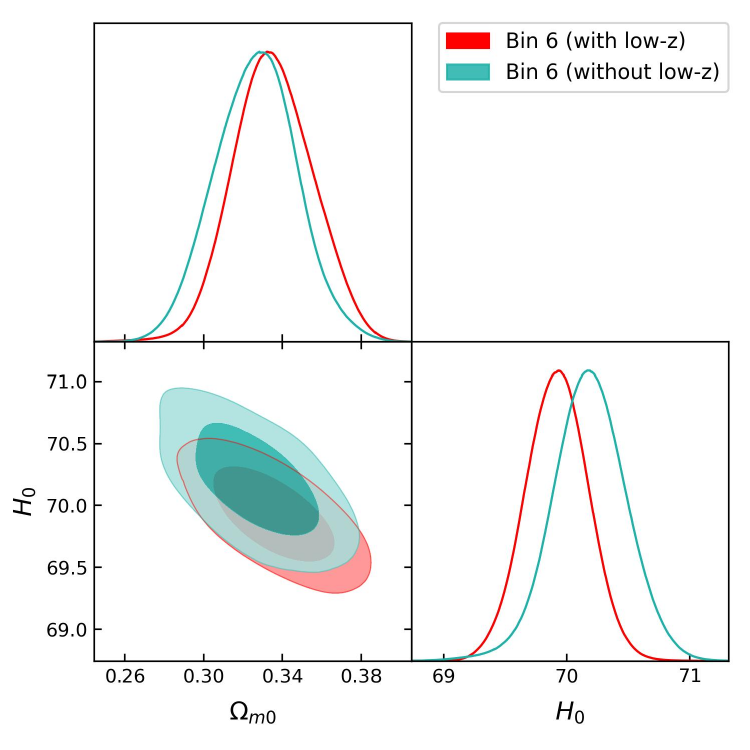}\hfill
    \includegraphics[width=0.22\textwidth]{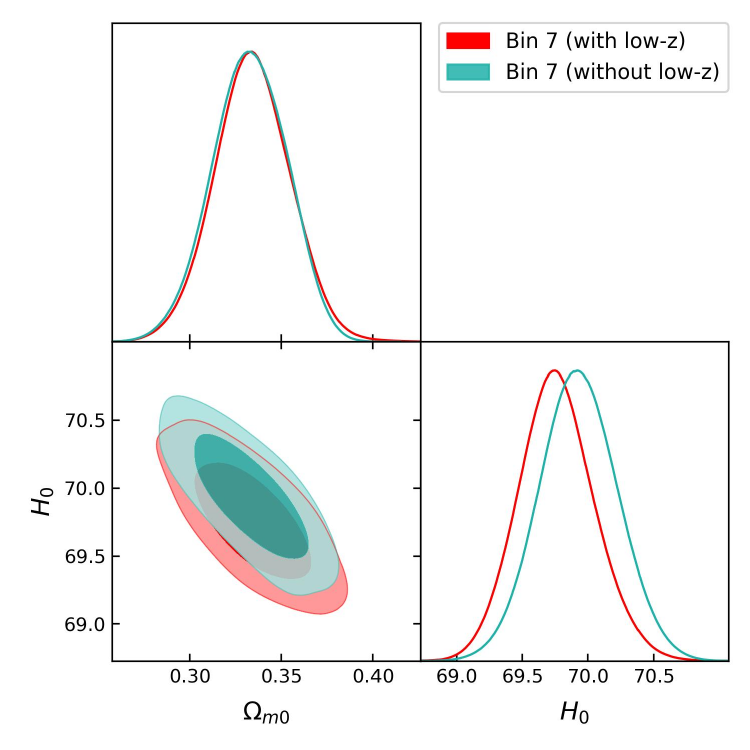}\hfill
    \includegraphics[width=0.22\textwidth]{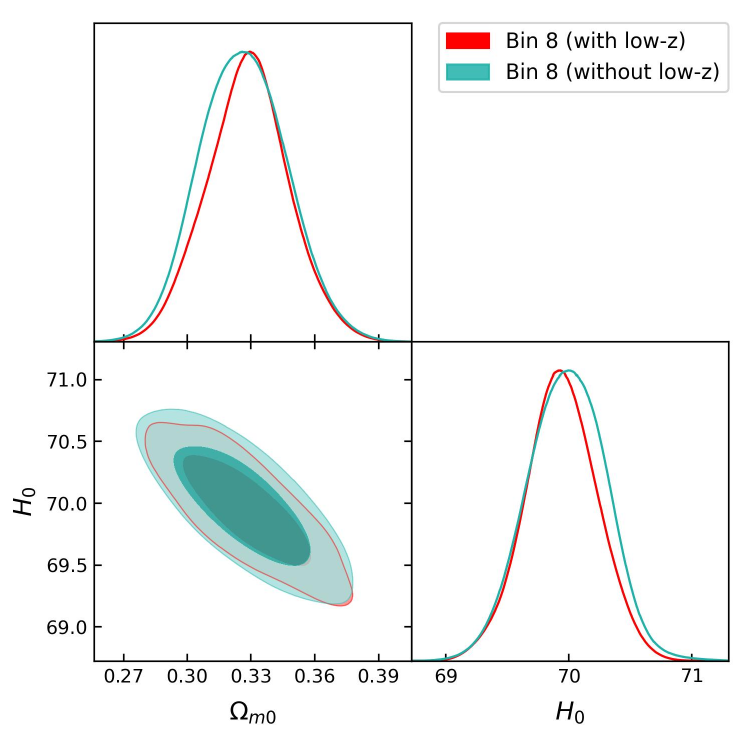}


    \includegraphics[width=0.22\textwidth]{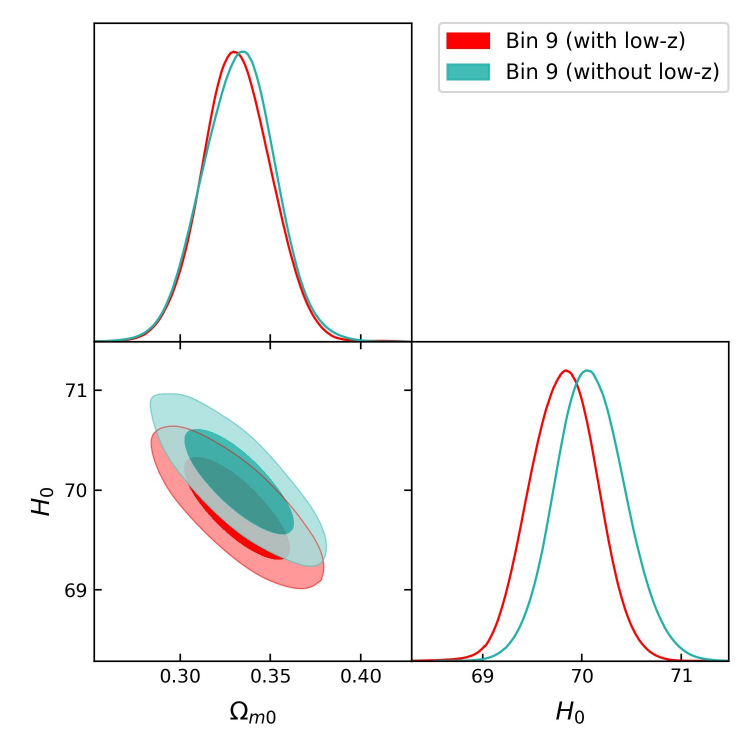}\hfill
    \includegraphics[width=0.22\textwidth]{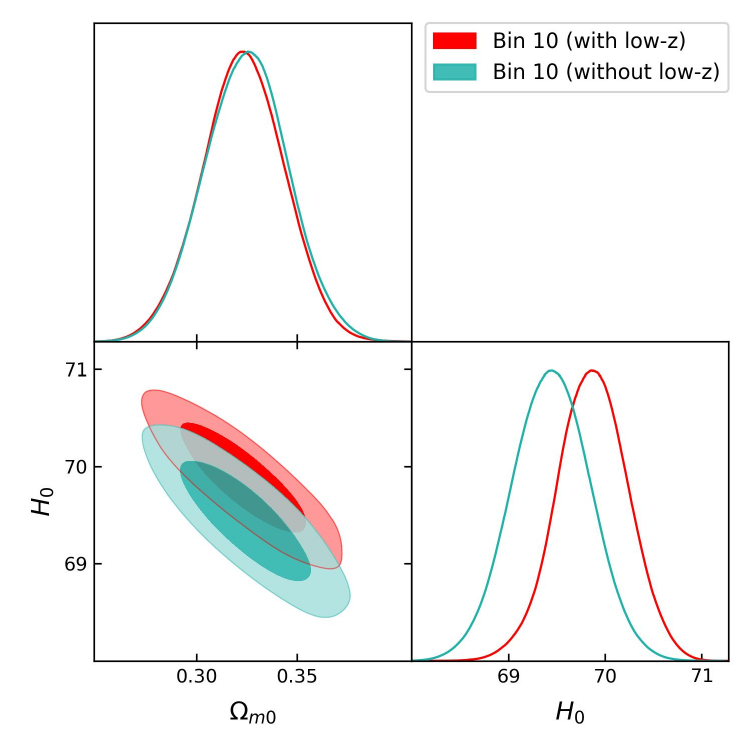}\hfill
    \includegraphics[width=0.22\textwidth]{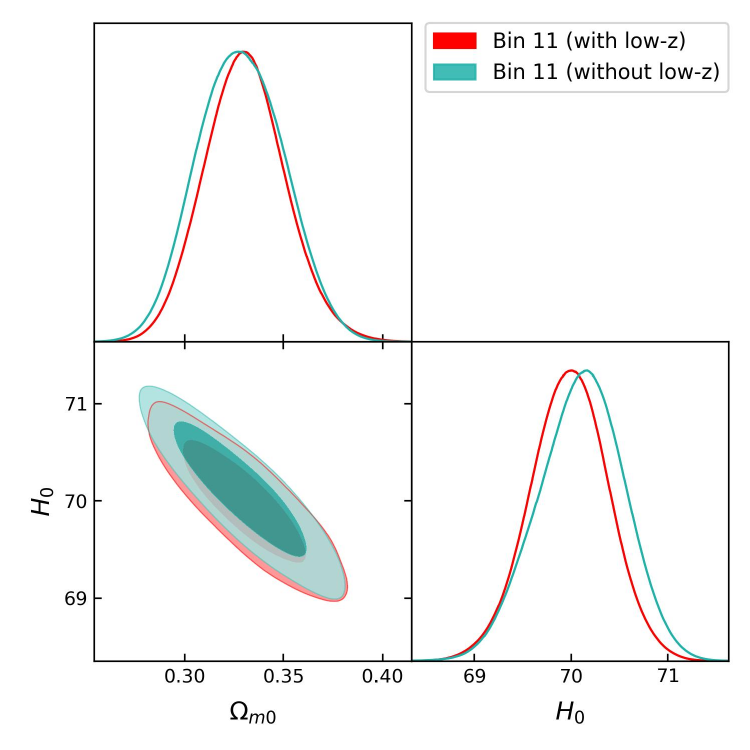}\hfill
    \includegraphics[width=0.22\textwidth]{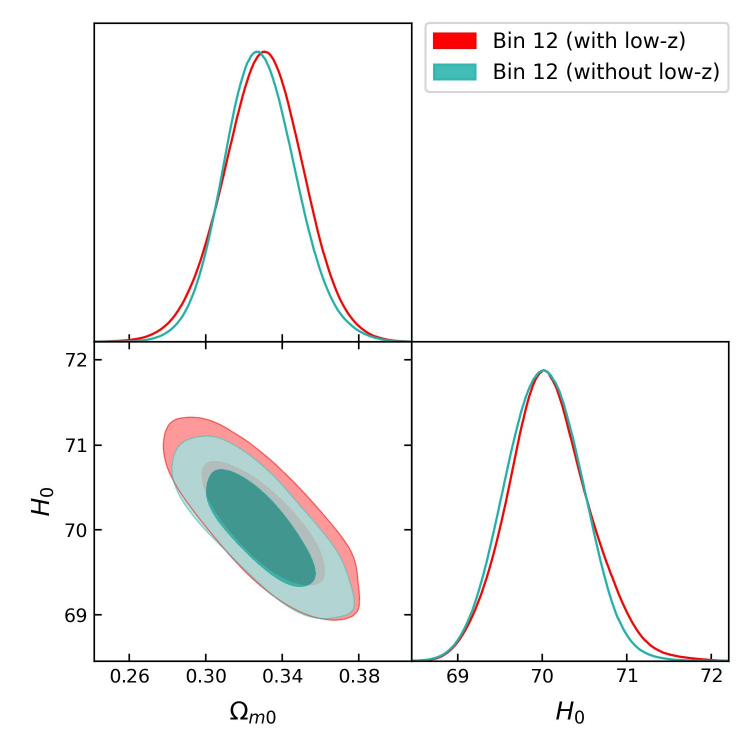}


    \includegraphics[width=0.22\textwidth]{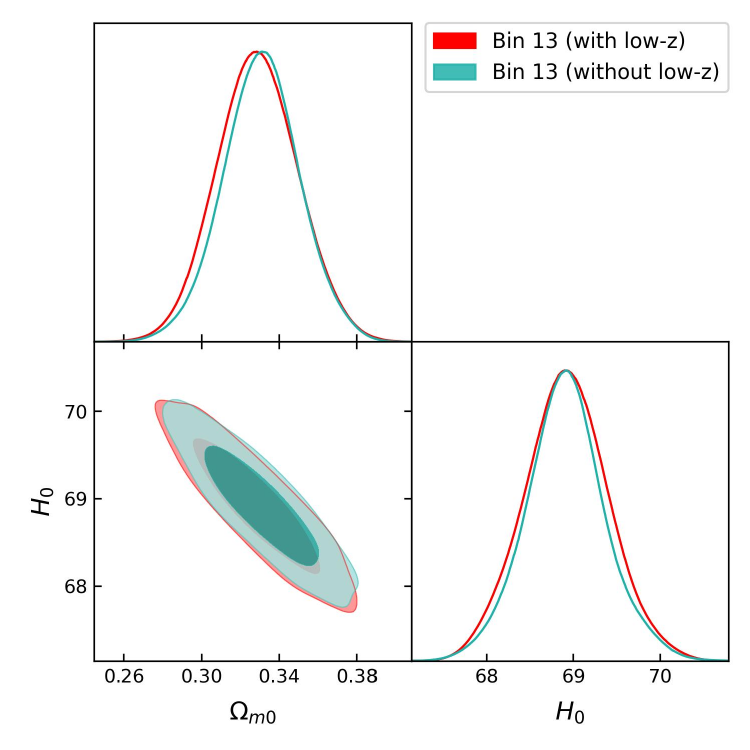}\hfill
    \includegraphics[width=0.22\textwidth]{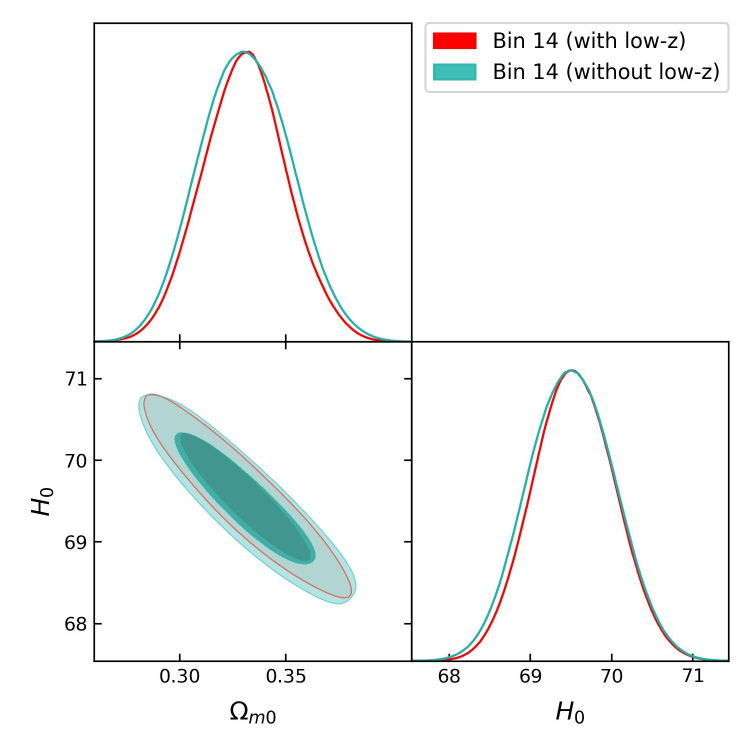}\hfill
    \includegraphics[width=0.22\textwidth]{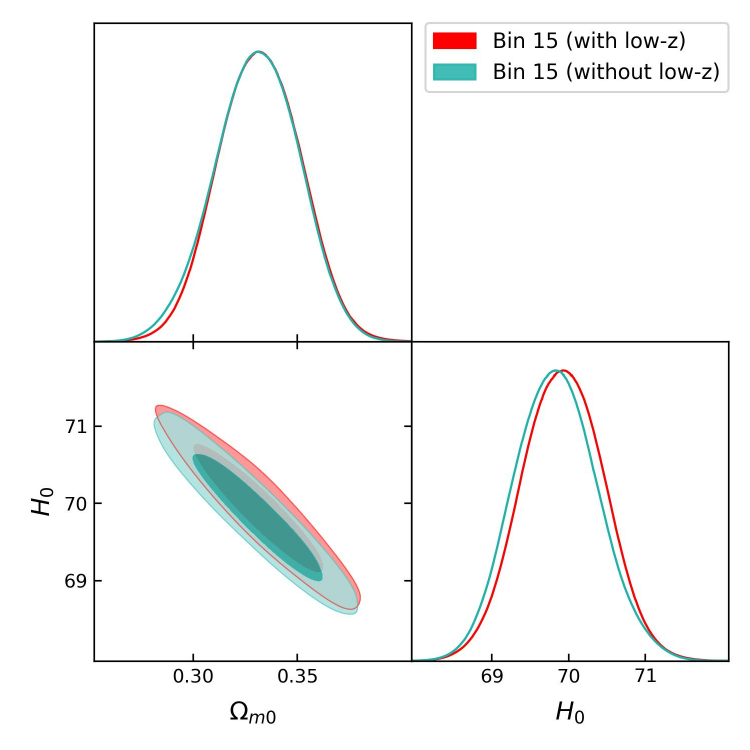}\hfill
    \includegraphics[width=0.22\textwidth]{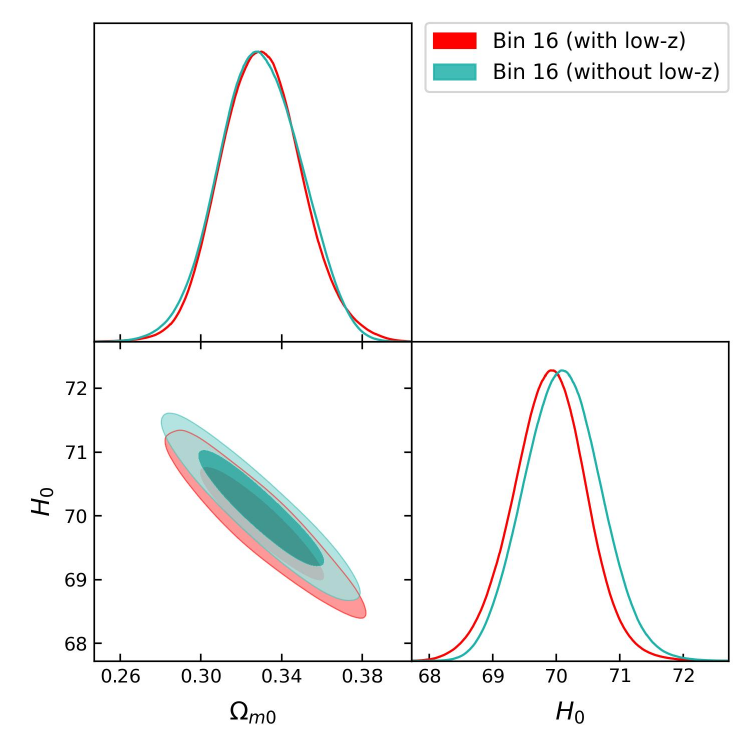}


    \includegraphics[width=0.22\textwidth]{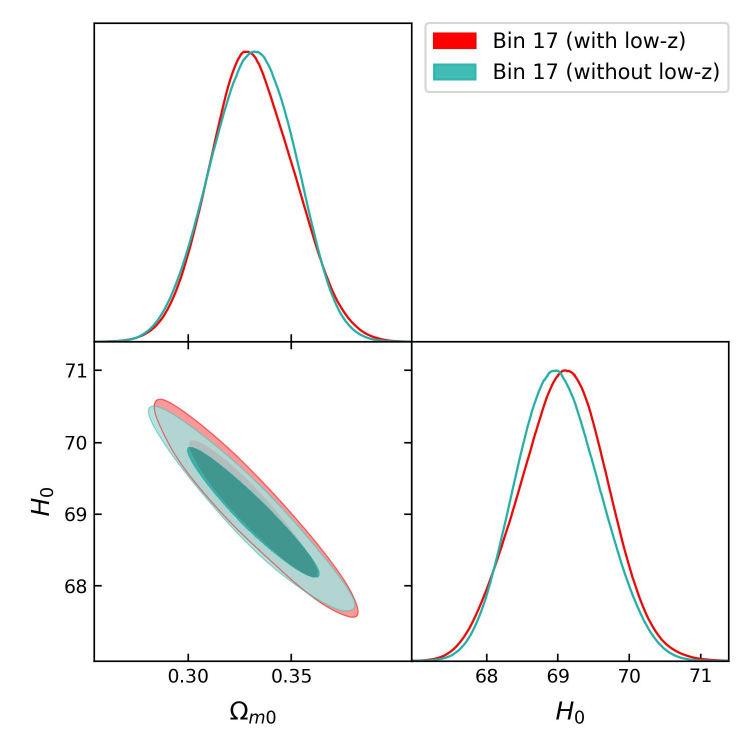}\hfill
    \includegraphics[width=0.22\textwidth]{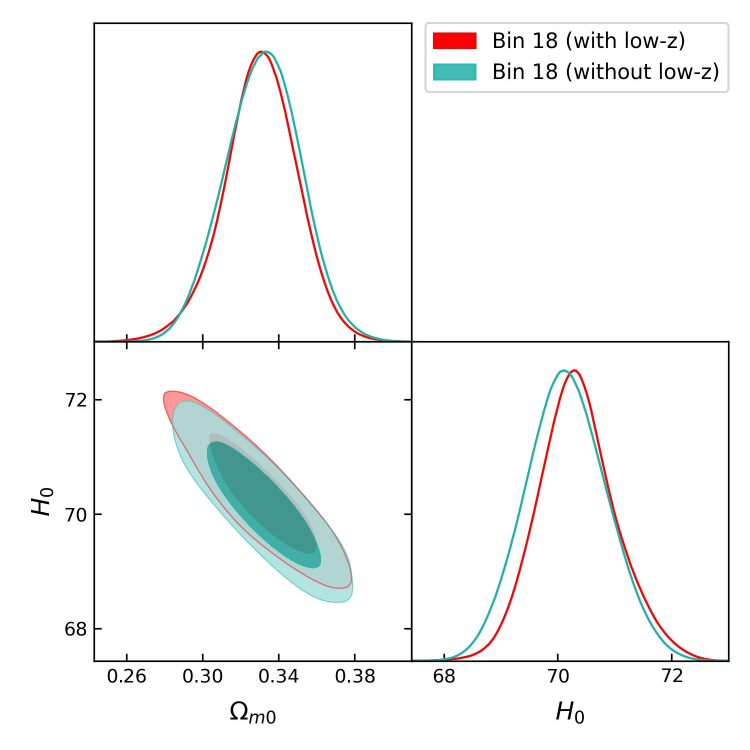}\hfill
    \includegraphics[width=0.22\textwidth]{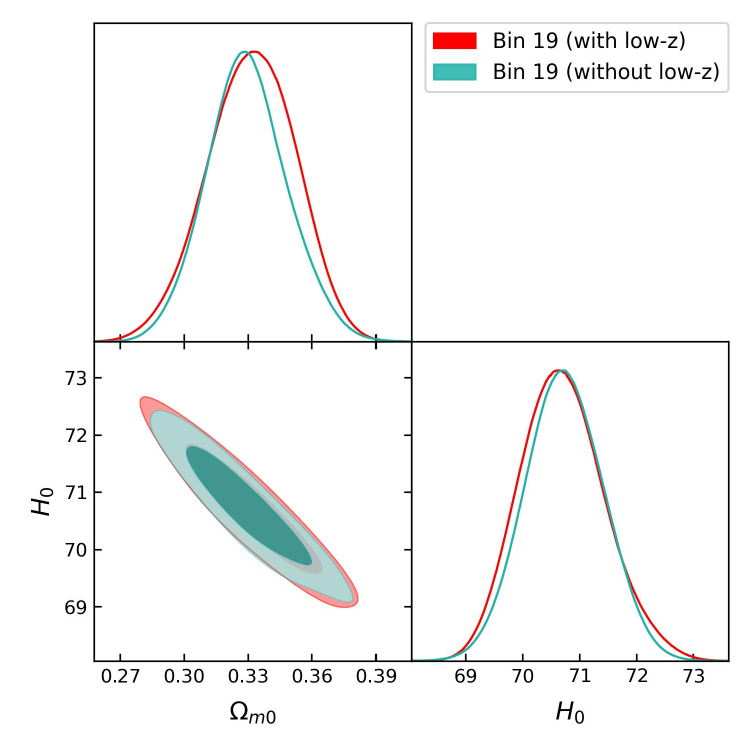}\hfill
    \includegraphics[width=0.22\textwidth]{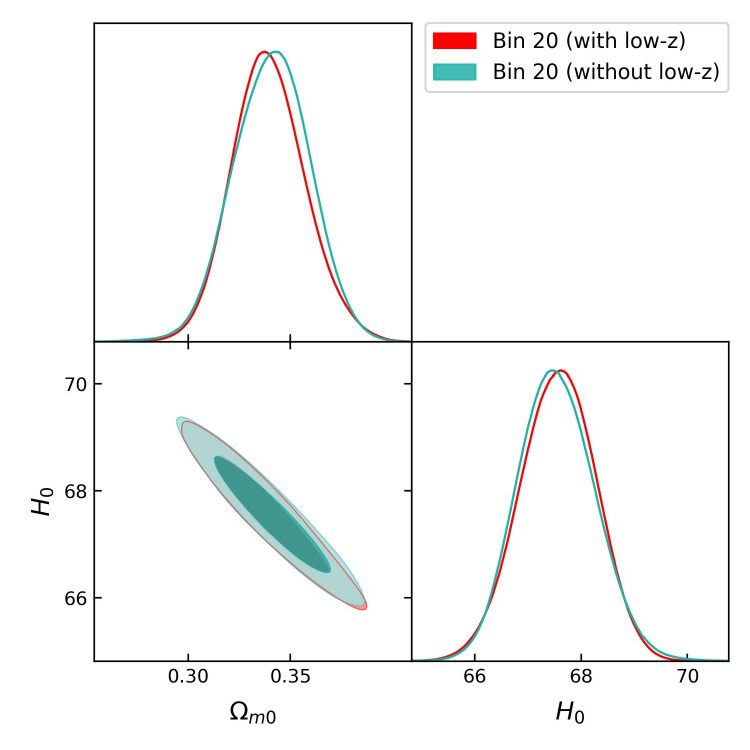}

    \caption{
    Joint two-dimensional posterior distributions of the cosmological parameters $H_{0}$ and
$\Omega_{\mathrm{m}0}$ obtained from the 20 redshift-bin analysis within the $\Lambda$CDM
model. Red contours show results for the Master Sample including low-$z$ supernovae, and light-seagreen contours correspond to the sample without low-$z$ data. The inner and outer contours denote the $1\sigma$ (68\%) and $2\sigma$ (95\%)
confidence regions, respectively.
}

    \label{fig:20panel}
\end{figure}

As mentioned in \autoref{sec:Results}, we have performed the data analysis for each bin, see \autoref{fig:20panel} where the joint posterior contour plots for each redshift are shown both for the Master Sample including the low-$z$ data (red contours) and for the one without low-$z$ values (green contours), at both the 68\% level (the darker contours) and the 95\% level (the lighter contours). As is visible from the figure, all bins are compatible at the 68\% level, with the exception of bin 3, which is compatible at the 95\% level.

\section*{Acknowledgements}
We thank Kazunori Kohri for the interesting discussion held during the Cosmology seminars at NAOJ and for the suggestion on the extrapolation to the BBN. A.B. acknowledges the academic support of the Department of Physics, The University of Burdwan, during the course of this study and the support of the Exploratory Research Fund from the Division of Science at NAOJ. 
M.G.D. acknowledges the support of the JSPS Grant-in-Aid for Scientific Research (KAKENHI) (A), Grant Number JP25H00675.


\bibliography{references}{}
\bibliographystyle{aasjournalv7}



\end{document}